\documentclass[lettersize,journal]{IEEEtran}
\usepackage{graphicx}
\usepackage[caption=false,font=normalsize,labelfont=sf,textfont=sf]{subfig}
\usepackage{amsthm,amsmath,amssymb}
\usepackage{ylqsymbol}

\usepackage{mathrsfs}
\usepackage{stfloats}
\usepackage{algorithm}
\usepackage{algpseudocode}
\usepackage{bm}
\usepackage{color}
\usepackage{comment}
\usepackage{caption}
\usepackage{cite}
\usepackage{leftidx}
\usepackage{epstopdf}
\usepackage{multirow}
\usepackage{booktabs} 

\begin{document}
\captionsetup[figure]{labelfont={default},labelformat={default},labelsep=period,name={Fig.},labelsep=period}

\title{Synesthesia of Machines (SoM)-Based Task-Driven MIMO System for Image Transmission}

\author{{Sijiang Li, \IEEEmembership{Graduate Student Member, IEEE}, Rongqing Zhang, \IEEEmembership{Senior Member, IEEE}, Xiang Cheng, \IEEEmembership{Fellow, IEEE} and Jian Tang,~\IEEEmembership{Fellow,~IEEE}}

\thanks{Manuscript received 24 April 2025; revised 20 July 2025; accepted 26 August 2025. This work was supported in part by the by the National Natural Science Foundation of China under Grant
62125101, Grant 62341101, and Grant 62271351; in part by the New Cornerstone Science Foundation through the XPLORER PRIZE. (\textit{Corresponding author: Xiang Cheng.})}
\thanks{Sijiang Li and Xiang Cheng are with the State Key Laboratory of Photonics and Communications,
School of Electronics, Peking University, Beijing, 100871, P. R. China (email: \{pkulsj, xiangcheng\}@pku.edu.cn).}
\thanks{Rongqing Zhang is with Intelligent Transportation Thrust, The Hong Kong University of Science and Technology (Guangzhou), Guangzhou, China (e-mail: rongqingz@tongji.edu.cn).}
\thanks{Jian Tang is with Beijing Innovation
Center of Humanoid Robotics, Beijing 101111, P. R. China (e-mail: jian.tang@x-humanoid.com).}
}

\markboth{}%IEEE Transactions on Wireless Communications,~Vol.~14, No.~8, August~2015
{Zeng \MakeLowercase{\textit{et al.}}: Bare Demo of IEEEtran.cls for IEEE Journals}

\maketitle
\vspace{-10mm}
\begin{abstract}
To support cooperative perception (CP) of networked mobile agents in dynamic scenarios, the efficient and robust transmission of sensory data is a critical challenge.
Deep learning-based joint source-channel coding (JSCC) has demonstrated promising results for image transmission under adverse channel conditions, outperforming traditional rule-based codecs. 
While recent works have explored to combine JSCC with the widely adopted multiple-input multiple-output (MIMO) technology, these approaches are still limited to the discrete-time analog transmission (DTAT) model and simple tasks.
Given the limited performance of existing MIMO JSCC schemes in supporting complex CP tasks for networked mobile agents with digital MIMO communication systems, this paper presents a Synesthesia of Machines (SoM)-based task-driven MIMO system for image transmission, referred to as SoM-MIMO.
By leveraging the structural properties of the feature pyramid for perceptual tasks and the channel properties of the closed-loop MIMO communication system, SoM-MIMO enables efficient and robust digital MIMO transmission of images. 
Experimental results have shown that compared with two JSCC baseline schemes, our approach achieves average mAP improvements of 6.30 and 10.48 across all SNR levels, while maintaining identical communication overhead.
\end{abstract}


\IEEEpeerreviewmaketitle

\section{Introduction}

\IEEEPARstart{I}{n} the era of beyond fifth generation (B5G) and sixth generation (6G), a large number of mobile agents, including autonomous vehicles, unmanned aerial vehicles, and humanoid robots, etc., will interact in real-time and execute diverse intelligent functions, revolutionizing industries and daily life. To enable diverse intelligent functionalities, such as decision-making and task execution, accurate environmental perception—encompassing the acquisition of object position, size, and category—is essential. Among various types of sensory data for perception, RGB images are widely used due to their rich semantic information, such as object color and texture. Additionally, they can be easily captured using cost-effective cameras. In the field of computer vision, RGB images have been extensively explored for perceptual tasks, including classification~\cite{classification}, object detection~\cite{yolo}, and instance segmentation~\cite{he2017mask}. However, in complex dynamic scenarios such as intelligent transportation systems, the quality of environmental perception is often compromised by adverse lightning conditions, harsh weather, and occlusions, leading to issues of missed detections and false positives. Benefiting from B5G/6G networks, multiple mobile agents can share perceptual information, enabling enhanced accuracy and beyond-line-of-sight perception, known as collaborative perception (CP)~\cite{gridmap,where2com,confidence}. Due to the latency and reliability requirements of perception in dynamic environments, high-speed and robust sensory data transmission emerges as a critical challenge in CP.

In typical digital communication systems for image transmission, source coding schemes, such as JPEG, JPEG2000, and BPG, are employed for image data compression, while channel coding schemes, including low-density parity-check (LDPC)~\cite{ldpc} and Turbo codes, are employed to correct bit errors in transmission. However, recent studies have shown that under adverse channel conditions, such traditional rule-based coding schemes suffer from rapid decoding performance degradation or even failure, which is known as the ``cliff effect"~\cite{DJSCC}. This issue makes it difficult for mobile agents to achieve efficient and reliable image transmission in low signal-to-noise ratio (SNR) scenarios, such as long distances, strong interference, and limited transmission power. 
To enhance image transmission performance under such conditions, the deep learning-based joint source-channel coding (JSCC) schemes have been proposed. By transmitting features extracted by ANNs, higher transmission efficiency and information fidelity can be achieved under adverse channel conditions, compared to traditional rule-based codecs.

Some recent works on JSCC have focused on combining JSCC with multiple-input multiple-output (MIMO) technology, which have been widely adopted in communication systems since the era of fourth generation (4G). However, existing MIMO JSCC schemes for image transmission still subject to several notable limitations. Firstly, these works rely on the discrete-time analog transmission (DTAT) model, assuming that the continuous features output by the ANNs can be directly transmitted over the channel, without addressing their integration with digital MIMO communication systems. Secondly, these works concentrate on optimizing image peak signal-to-noise ratio (PSNR) and structural similarity (SSIM) for image reconstruction tasks, failing to consider the performance of typical downstream perceptual tasks of mobile agents. 
These limitations, which arise from the lack of deep coupling between the digital MIMO communication system and the perceptual tasks, substantially constrain the potential of MIMO JSCC schemes in empowering CP for mobile agents.
Inspired by human synesthesia, in which the stimulation of one sense organ will automatically evoke another sense organ to jointly perform cognitive tasks, the Synesthesia of Machines (SoM) framework is proposed in~\cite{SoM}. The SoM framework refers to the intelligent integration of multi-modal information from disparate ``machine senses", such as communication devices and sensors, to perform intelligent functionalities more effectively. Following this principle, artificial neural networks (ANNs) can be utilized to jointly encode MIMO channel information from communication devices and image information from sensors into compact, task-aware and robust features, referred to as SoM-features, constructing a low-latency, high-reliability digital MIMO transmission scheme.

In this paper, we propose SoM-MIMO, an SoM-based task-driven MIMO system for image transmission. By leveraging the structural properties of the feature pyramid for perceptual tasks and the channel properties of the closed-loop MIMO communication system, SoM-MIMO enables efficient and robust digital MIMO transmission of images. 
To demonstrate the performance of the proposed scheme, we employ the challenging instance segmentation task in complex scenarios. Compared with two JSCC baseline schemes, our approach achieves average mAP improvements of 6.30 and 10.48 across all SNR levels, while maintaining identical communication overhead\footnote{Simulation codes are provided to reproduce the results presented in this paper: https://github.com/SijiangLi/SoM-MIMO}. The contributions of this paper are in the following aspects:
\begin{enumerate}
\item To address the limitations of existing MIMO JSCC schemes in supporting complex perceptual tasks with digital communication systems, we propose the SoM-MIMO scheme, which enables efficient task-driven digital MIMO transmission of images.
\item To ensure efficient and robust task-driven image transmission, we design a hierarchical feature fusion module integrated with an MIMO channel-aware encoding module, which optimizes both the density and fidelity of task-critical information during its transmission over MIMO fading channels. 
\item To enhance the adaptability to digital communication systems, we integrate nonlinear feature activation into standard digital MIMO baseband processing, which improves the generality and reliability of our scheme in digital transmission systems.
\end{enumerate}

The remainder of this paper is structured as follows. The related work on JSCC are presented in Sec.~\ref{sec:related}. The system model is presented in Sec.~\ref{sec:sys}. The proposed SoM-MIMO scheme is presented in Sec.~\ref{sec:frame}. Then the experimental results are given in Sec.~\ref{sec:simu} to evaluate the performance of the proposed scheme. Finally, conclusions and ongoing research issues are highlighted in Sec.~\ref{sec:conclusion}.

\section{Related Work}\label{sec:related}
An early JSCC scheme based on convolution neural network (CNN) was proposed in~\cite{DJSCC}. Through end-to-end optimization on the PSNR of the image reconstruction, it achieved better performance than traditional schemes under both additive white Gaussian noise (AWGN) and Rayleigh fading channels. 
To enhance the adaptability of JSCC scheme to varying channel conditions,~\cite{xu2021wireless} explicitly incorporated SNR information through feature attention modules. Furthermore, \cite{witt} employed the Swin Transformer blocks to enhance the feature representation capability of the proposed JSCC scheme. The authors in \cite{fast} further considered time-varying channel conditions and allocated essential features to high-quality channels based on channel prediction results. 

While the aforementioned works focus on single-input single-output (SISO) scenarios, \cite{early-mimo, openloop-zp, openloop-csg, closeloop-ab, closeloop-scan, closeloop-deniz, closeloop-diffusion} have extended JSCC to MIMO systems. 
Early work in~\cite{early-mimo} explored a CNN-based JSCC scheme for bit transmission in MIMO systems. 
For image transmission tasks, the authors in \cite{openloop-zp} proposed an MIMO spatial multiplexing mechanism with an adaptive coding rate, which utilized channel quality indicators (CQI) feedback to the transmitter. 
While for closed-loop MIMO systems, where the transmitter has CSI information, a non-invasive CSI embedding scheme was designed in~\cite{openloop-csg}. 
Without explicitly using traditional MIMO precoding, this scheme leveraged CSI information to generate a mask applied to the multi-head attention weights in the Swin Transformer blocks, thereby enabling adaptivity to MIMO channel conditions. 
However, more recent works have adopted explicit MIMO precoding schemes, such as singular value decomposition (SVD) precoding. 
\cite{closeloop-ab} and \cite{closeloop-scan} utilized the equivalent SNR obtained from SVD precoding and the full CSI information, respectively, to achieve feature adaptive encoding based on attention modules. 
While \cite{closeloop-deniz} fed the equivalent SNR and the original image into ViT modules, leveraging the self-attention mechanism to achieve JSCC in MIMO systems with SVD precoding. 
The diffusion model was also utilized in \cite{closeloop-diffusion} to perform feature denoising based on the feature distribution at the receiver.

To bridge the gap between JSCC and digital communication systems, \cite{digital-deniz} and \cite{digital-tmx} designed joint coding-modulation schemes, where features were mapped to digital symbols in a learnable manner. By decoupling quantization and modulation, \cite{digital-ab,sdac} designed learnable nonlinear quantization mechanisms to achieve the conversion of continuous features to bits. As for downstream perceptual tasks, \cite{digital-tmx,digital-ab,task-class} considered simple image classification and retrieval tasks, while \cite{task-segmentation} addressed the semantic segmentation task. However, MIMO systems and channel conditions were not involved in \cite{task-segmentation}. Moreover, no existing JSCC scheme has addressed the more challenging tasks for mobile agents, such as instance segmentation using high-resolution images in complex environments.
As summarized in Table~\ref{tab:existing_work}, existing schemes have not simultaneously addressed the demands of digital MIMO systems and complex perceptual tasks, which makes them insufficient to support CP tasks for networked intelligent agents.
In contrast, our SoM-MIMO scheme is fundamentally task-driven, directly optimizing performance for the challenging perceptual task within a practical digital MIMO framework that utilizes SVD precoding.

\begin{table*}[!t]
\centering
{
\caption{Summary of existing work related to SoM-MIMO.} 
\begin{tabular}{|c|c|c|c|c|}
\hline
\textbf{Existing work} & \textbf{System configuration} & \textbf{Transmission scheme} & \textbf{Task} & \textbf{Channel embedding}\\
\hline
\cite{DJSCC} & SISO & DTAT & Reconstruction & None\\
\hline
\cite{xu2021wireless, witt} & SISO & DTAT & Reconstruction & SNR \\
\hline
\cite{fast} & SISO & DTAT & Reconstruction & CSI \\
\hline
\cite{task-class} & SISO & DTAT & Retrieval & None \\
\hline
\cite{task-segmentation} & SISO & DTAT & Semantic segmentation & None \\
\hline
\cite{digital-deniz, sdac} & SISO & Digital & Reconstruction & None \\
\hline
\cite{digital-tmx} & SISO & Digital & Reconstruction \& Classification  & None \\
\hline
\cite{digital-ab} & SISO & Digital & Classification  & None \\
\hline
\cite{openloop-zp} & MIMO w/o precoding & DTAT & Reconstruction  & CQI \\
\hline
\cite{openloop-csg} & MIMO w/o precoding & DTAT & Reconstruction  & CSI \\
\hline
\cite{closeloop-ab,closeloop-deniz,closeloop-diffusion} & MIMO with SVD & DTAT & Reconstruction  & Equivalent SNR \\
\hline
\cite{closeloop-scan} & MIMO with SVD & DTAT & Reconstruction  & CSI, noise power \\
\hline
\cite{early-mimo} & MIMO with SVD & Digital & Bit transmission  & Equivalent SNR \\
\hline
\textbf{Ours} & \textbf{MIMO with SVD} & \textbf{Digital} & \textbf{Instance segmentation}  & \textbf{Equivalent SNR} \\
\hline
\end{tabular}
\label{tab:existing_work}
}
\vspace{-4mm}
\end{table*}

\section{System Model}\label{sec:sys}
\subsection{Image Instance Segmentation}\label{sec:seg}
\begin{figure}[!t]
\centering{}
\includegraphics[width=0.50\textwidth]{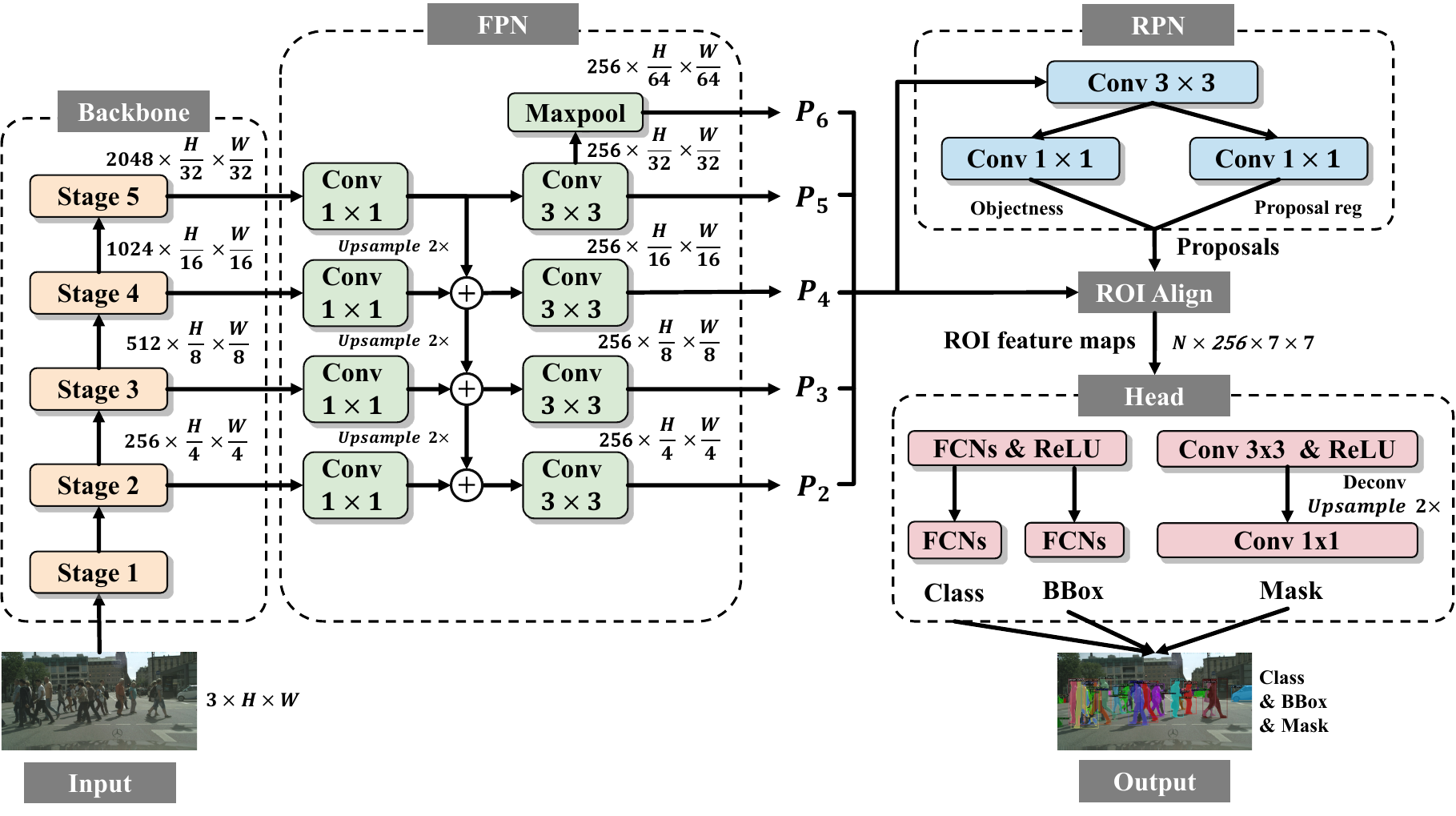}
\caption{Diagram of Mask R-CNN with FPN.}{}
\label{MaskRCNN}
\vspace{-6mm}
\end{figure}

Instance segmentation is widely acknowledged as a challenging task in computer vision. Different from object detection, where the goal is to classify objects in the image and localize them with bounding boxes, and semantic segmentation, where the goal is to classify each pixel in the image into different classes without distinguishing between different instances, it requires not only locating and classifying each objects but also generating pixel-level masks for them.

Mask R-CNN~\cite{he2017mask} is regarded as a conceptually simple, flexible, and general framework for image instance segmentation. By extending Faster R-CNN~\cite{ren2016faster} with an additional FCN branch to predict segmentation masks, it enables multi-task learning in a flexible manner. Feature Pyramid Network (FPN)~\cite{lin2017feature} is a commonly adopted approach in various state-of-the-art computer vision models for enhancing multi-scale feature representation. Considering feature maps with different spatial resolution can be extracted by feature backbone, such as ResNet~\cite{he2016deep} and Swin Transformer~\cite{liu2021swin}, FPN combines high-level, low-resolution feature maps with low-level, high-resolution feature maps to improve detection performance for small-scale and multi-scale objects. A typical Mask R-CNN structure with FPN is shown in Fig.~\ref{MaskRCNN}. An RGB image $\bbI \in \mathbb{R}^{3 \times H \times W}$ is fed into the backbone and FPN to generate the pyramid feature maps $\bbP_{2}, \bbP_{3}, \dots, \bbP_{6}$. These feature maps are then passed through the Region Proposal Network (RPN), followed by Region of Interest (ROI) Align, and finally processed by the multi-task head for instance segmentation. It is evident that the pyramid feature maps contain rich information required for instance segmentation tasks. However, even without considering the bit differences between floating-point and integer representations, the data volume of these pyramid feature maps can still be more than seven times that of the image. Therefore, directly transmitting them is impractical. Nevertheless, their sparsity makes further encoding possible.

\subsection{Closed-loop MIMO Precoding}\label{seg:precoding}
For a fully digital $N \times N$ MIMO systems, the symbols to be transmitted over the channel are denoted as $\bbX \in \mathbb{C}^{N \times k}$, where $k$ denotes the number of MIMO symbols. For each MIMO symbol $\bbX_{i}$, the power $\bbE_{s}$ is constrained as:
\begin{equation}
\begin{aligned}
E_s=||\bbX_{i}||^2_2 = P_s,
\end{aligned}
\label{power}
\end{equation}
where $||.||_2$ denotes the Euclidean Norm and $i=1,2,\dots,k$. We can set $P_s=1$ without loss of generality.

The channel model can be formulated as:
\begin{equation}
\begin{aligned}
\bbY=\bbH\bbX+\bbN,
\end{aligned}
\label{channel}
\end{equation}
where $\bbY\in\mathbb{C}^{N \times k}$ denotes the channel output symbols. For a Rayleigh fading channel, each element $\bbH[i,j]$ in the channel matrix $\bbH \in \mathbb{C}^{N \times N}$ follows a complex Gaussian distribution with zero mean and variance $\sigma_h^{2}$. We perform power normalization to the channel matrix to maintain a consistent SNR at the transmitter and receiver, i.e.,
\begin{equation}
\begin{aligned}
\mathbb{E}[||\bbH\bbX_{i}||_{2}^{2}]=||\bbX_{i}||_{2}^{2}=E_s.
\end{aligned}
\label{channel_power}
\end{equation}
Correspondingly, the SNR can be defined as:
\begin{equation}
\begin{aligned}
\text{SNR}=10\log_{10}\frac{E_s}{E_N}=10\log_{10}\frac{E_s}{N\sigma_{n}^2}~\text{(dB)},
\end{aligned}
\label{SNR}
\end{equation}
where $\sigma_{n}^{2}$ denotes to the noise variance. We also consider the block-fading assumption, which means $\bbH$ remains constant throughout the transmission of features belong to one image.

In a closed-loop MIMO system, the CSI is available at both transmitter and receiver, enabling precoding at the transmitter and combining at the receiver. We utilize singular-value decomposition (SVD) precoding to decompose the channel matrix as $\bbH=\bbU\bm{\Sigma}\bbV^{\mathrm{H}}$, where $\bbU \in \mathbb{C}^{N \times N}$ and $\bbV \in \mathbb{C}^{N \times N}$ are unitary matrices, and $\bm{\Sigma}$ is a diagonal matrix consists of singular values in descending order as $\lambda_1 \ge \lambda_2 \ge \dots \ge \lambda_N$. At the transmitter, the precoding can be denoted as:
\begin{equation}
\begin{aligned}
\bbX=\bbV\bbX^{'},
\end{aligned}
\label{trans_precoding}
\end{equation}
where $\bbX^{'}$ denotes to the data symbols before precoding. At the receiver, symbols $\bbY^{'}$ can be denoted as:
\begin{equation}
\begin{aligned}
\bbY^{'}=\bbU^{\text{H}}\bbY=\bbU^{\text{H}}(\bbH\bbX+\bbN)=\bm{\Sigma}\bbX^{'}+\bbN^{'},
\end{aligned}
\label{recei_precoding}
\end{equation}
where $\bbN^{'}=\bbU^{\mathrm{H}}\bbN$ and follows the same distribution as $\bbN$ as $\bbU$ is a unitary matrix. Thus SVD precoding effectively decomposes the MIMO channel $\bbH$ into $N$ independent sub-channels, simplifying the system design. Each sub-channel's equivalent SNR can be expressed as:
\begin{equation}
\begin{aligned}
\text{SNR}_i=10\log_{10}\frac{\lambda_{i}^{2}E_s}{N\sigma_{n}^2}~\text{(dB)}.
\end{aligned}
\label{Eq_SNR}
\end{equation}
 
 At the receiver, channel equalization can be achieved using the singular value matrix $\bm{\Sigma}$ as:
 \begin{equation}
\begin{aligned}
\hat{\bbX}^{'}=\bm{\Sigma}^{-1}\bbY^{'}=\bbX^{'}+\bm{\Sigma}^{-1}\bbN^{'},
\end{aligned}
\label{equlization}
\end{equation}
followed by subsequent symbol detection.
\section{SoM-MIMO System Design}\label{sec:frame}
\begin{figure*}[!t]
\centering
\includegraphics[width=0.95\textwidth]{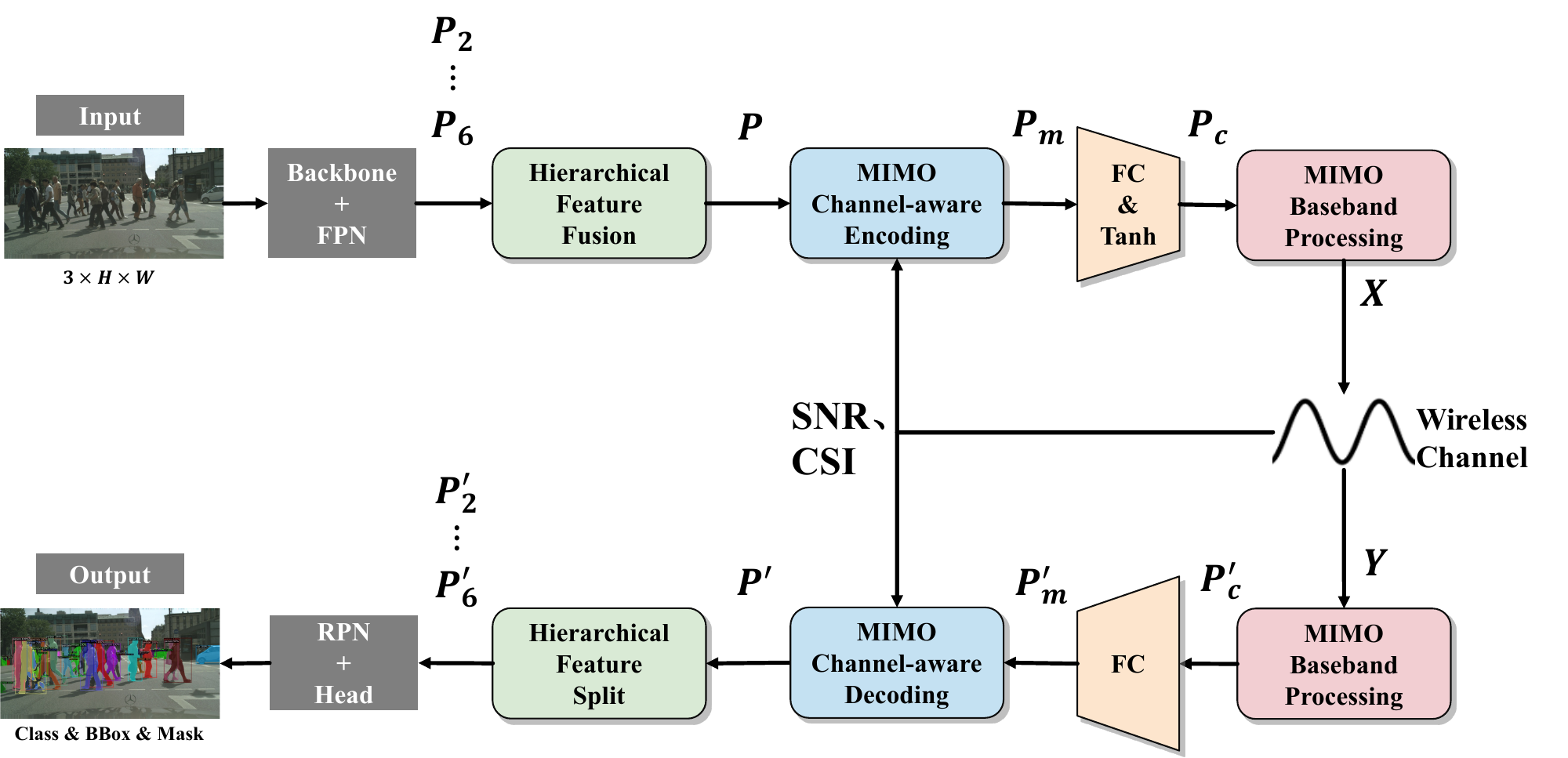}
\caption{Diagram of the proposed scheme.}
\label{Scheme}
\vspace{-6mm}
\end{figure*}
To achieve efficient and reliable cooperative perception in complex dynamic scenarios, we propose an SoM-MIMO scheme. The overall architecture of the scheme is illustrated in Fig.~\ref{Scheme}. At the transmitter, the feature pyramid extracted by the backbone and FPN is first fed into a hierarchical feature fusion (HFF) module to reduce information redundancy. Then, an MIMO channel-aware encoding (MCE) module is designed to utilize physical layer information, such as SNR and CSI, to enhance the robustness of feature transmission over fading MIMO channels. Subsequently, a fully connected (FC) layer and a nonlinear activation layer are employed for flexible feature channel compression and distribution transformation. Finally, standard MIMO baseband processing, including quantization, modulation, and precoding, is performed to obtain MIMO symbols $\bbX$, enabling our scheme to be natively compatible with current digital communication systems. Upon receiving $\bbY$, the receiver performs symmetric operations to reconstruct the feature pyramid, whose information fidelity is then validated through an instance segmentation task head. Modules at the receiver include MIMO baseband processing, an FC layer, an MIMO channel-aware decoding (MCD) module and a hierarchical feature split (HFS) module. The details of the proposed scheme are shown as follows.
\subsection{Hierarchical Feature Fusion and Split Modules}\label{sec:pyramid}
\begin{figure*}[ht]
    \centering
    \subfloat[Hierarchical Feature Fusion Module Diagram]{\includegraphics[width=0.5\textwidth]{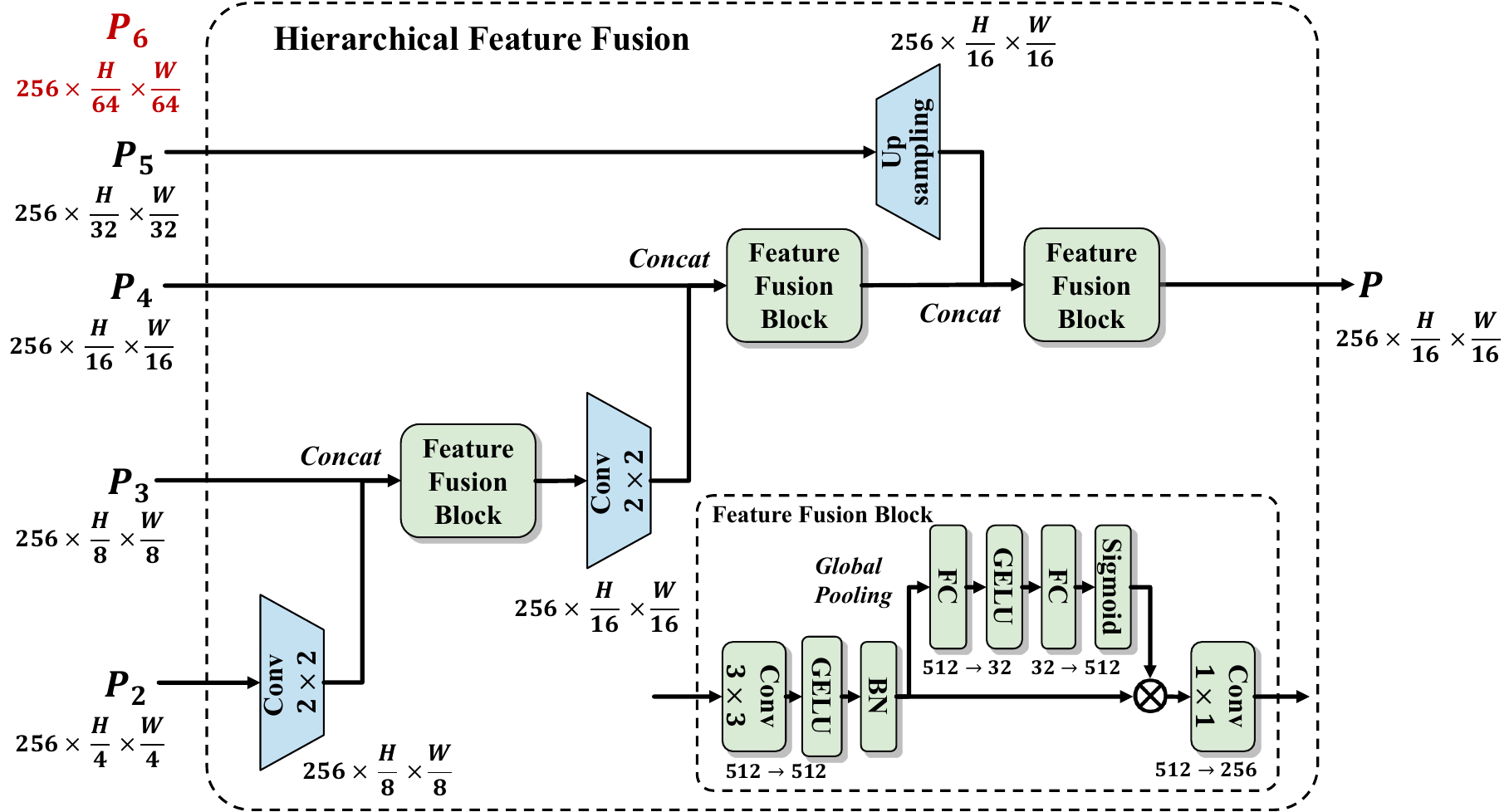}\label{fig:fusion}}
    \hfill
    \subfloat[Hierarchical Feature Split Module Diagram]{\includegraphics[width=0.5\textwidth]{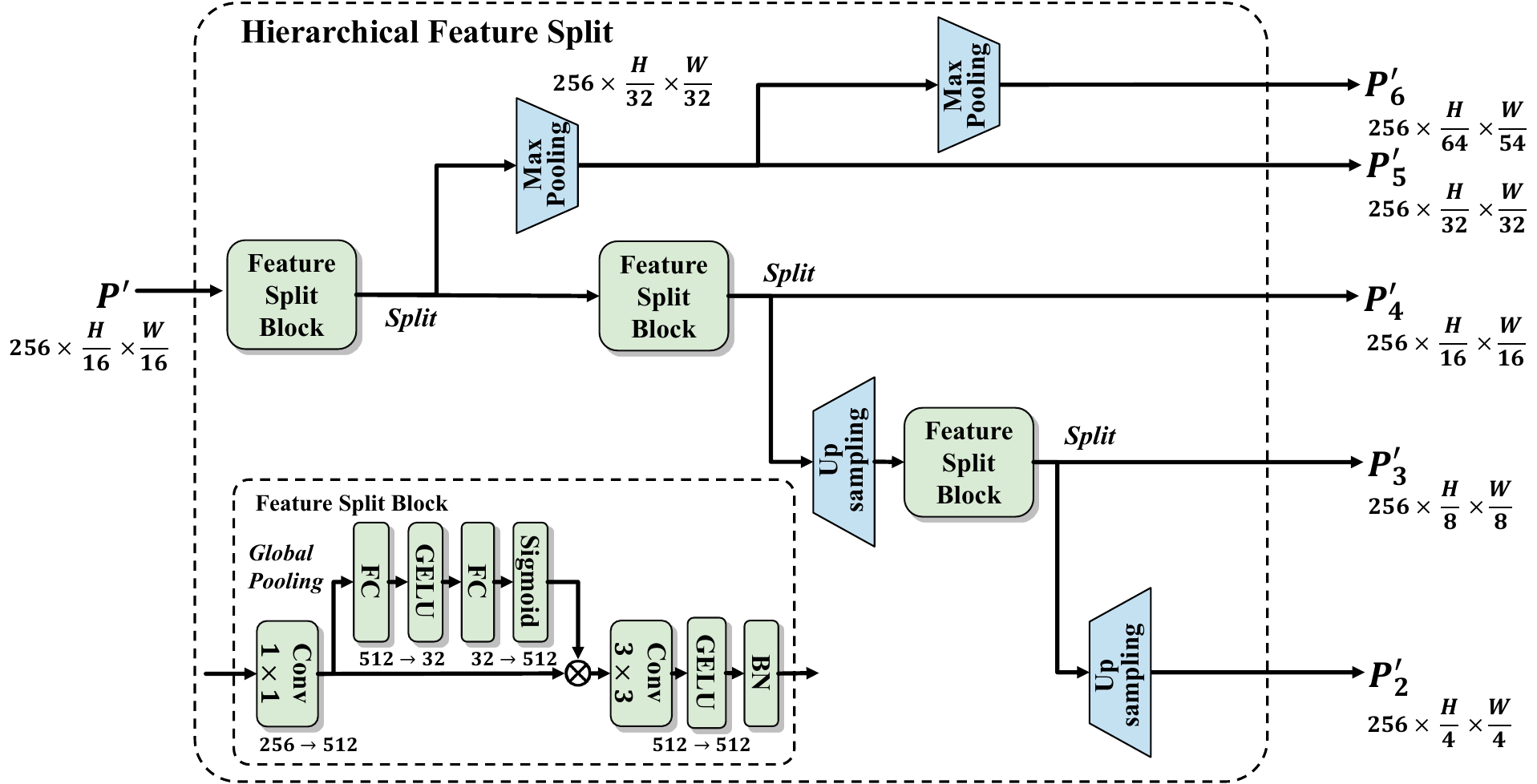}\label{fig:split}}
    \caption{Diagram of the hierarchical feature fusion and split modules for feature pyramid processing.}
    \label{fig:fusion and split}
\vspace{-4mm}
\end{figure*}

After acquiring the environmental image, the transmitter extracts the feature pyramid as mentioned in Sec.~\ref{sec:seg}. The feature pyramid can effectively enhance the feature representation and small object detection capabilities, but suffers from large data size, which is unfavorable for low-latency transmission. Considering that the feature pyramid represents multi-scale feature expression of the same image, the correlation of features can be utilized for fusion and compression. The detail of the modules deployed at the transmitter and receiver for feature pyramid fusion and split are shown in Fig.~\ref{fig:fusion and split}.

At the transmitter, starting from $\bbP_2$, a feature map with higher spatial resolution is first downsampled through a $2\times2$ convolutional Layer with $\text{stride}=2$, and then channel-wise concatenated with the feature map of lower resolution. Subsequently, the concatenated feature map that contains richer information is fed into a feature fusion block to achieve channel aggregation. 

In the feature fusion block, a $3\times3$ convolutional layer followed by a Gaussian error linear unit (GELU) and batch normalization (BN) is first utilized to fuse the information from the two concatenated feature maps as well as capturing spatial context. 
Then a squeeze and excitation (SE) block~\cite{senet} with GELU activation is added to provide channel-wise attention and enhance feature representation ability of the model.
Finally, the feature map ie fed into a $1\times1$ convolutional layer for channel compression. Compared to the input feature map, the output feature map reduces the number of channels by half while maintaining the same spatial resolution.

Following the sequence of $P_2$ to $P_5$, the fused feature is downsampled hierarchically and then concatenated with new original feature maps for further fusion, as shown in Fig.~\ref{fig:fusion}. The final resolution is constrained to be the same as $P_4$ to balance information fidelity with data volume, as a lower resolution would lead to the loss of detailed object information, while a higher resolution would introduce significant data redundancy for transmission.
Therefore, $P_5$ needs to be upsampled before fusion, and the upsampling method is bilinear interpolation. 
After passing through the feature fusion module, the feature pyramid $P_2$ to $P_6$ with large information redundancy (with a feature volume of $V_{in}= 256 \times \sum_{k=2}^{5} \left( \frac{H}{2^k} \times \frac{W}{2^k} \right) \approx 21.3HW$) has been fused into a feature map $P$ (with a feature volume of $V_{out}= 256 \times \frac{H}{16} \times \frac{W}{16}= HW$), which leads to a reduction of over 95\% in data volume.

At the receiver, the feature map $P^{'}$ will pass through an HFS module that is symmetric to the HFF module. Specifically, the feature fusion block at the transmitter is replaced with a feature split block, whose internal structure follows a mirrored sequence. For the output of the feature split block, the second half of the channels will serve as the output of the reconstructed feature pyramid after adjusting the spatial resolution, and the first half of the channels will be upsampled and fed into the subsequent feature separation block, as shown in Fig.~\ref{fig:split}. The upsampling and downsampling methods at the receiver are bilinear interpolation and max pooling respectively.
 
Note that, since $P_6$ is directly obtained by downsampling from $P_5$ in the previous FPN module, there is no need to feed $P_6$ into the feature fusion module. And at the receiver, $P^{'}_6$ can be obtained by performing max pooling on the reconstructed feature map $P^{'}_5$, respectively.

\subsection{MIMO Channel-aware Encoding and Decoding Modules}\label{sec:mimo}
\begin{figure}[!t]
\centering{}
\includegraphics[width=0.50\textwidth]{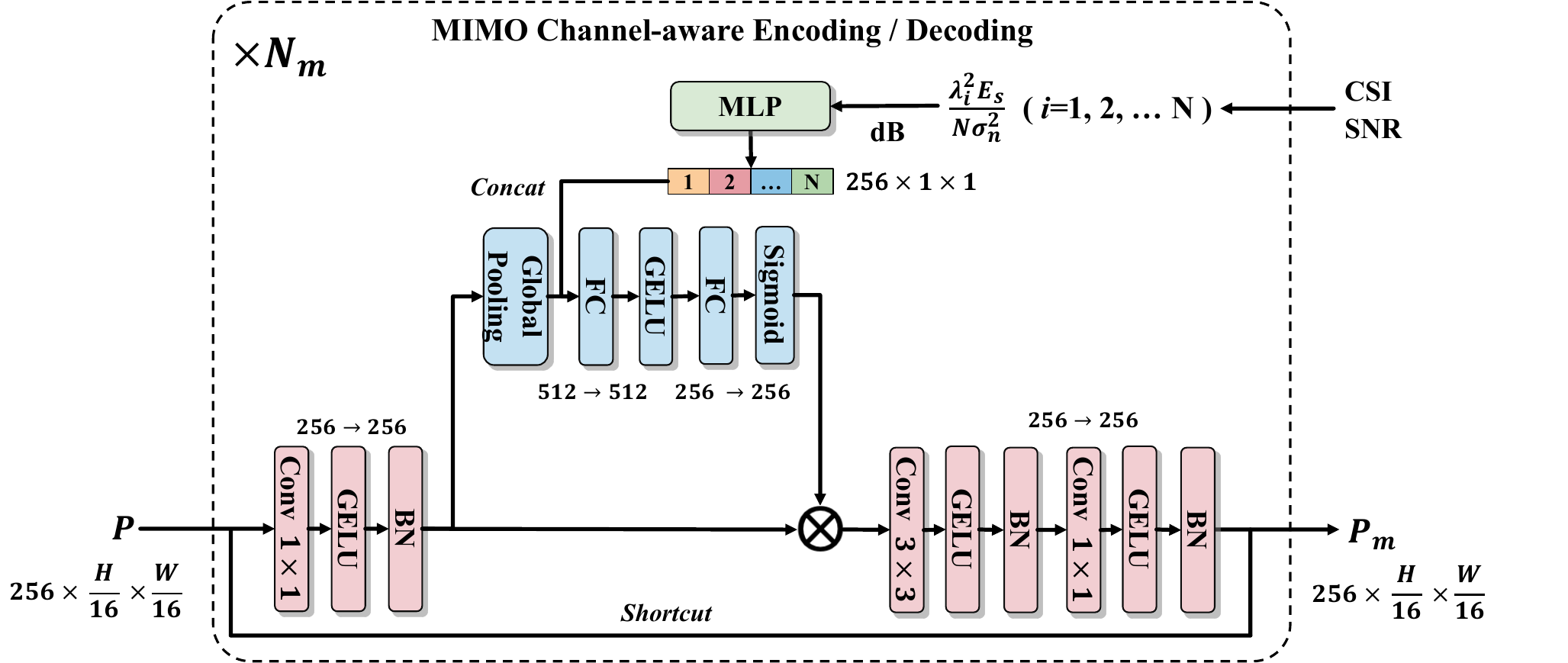}
\caption{Diagram of the MIMO channel-aware encoding and decoding modules.}{}
\label{MIMO_encoding}
\vspace{-6mm}
\end{figure}
After fusing the feature pyramid into $P$, subsequent feature processing can be performed by incorporating MIMO channel information. Considering the precoding method in Sec.~\ref{seg:precoding}, data transmitted over different sub-channels may suffer from different levels of distortion, due to varying equivalent SNR. 
In the feature map $P$, each 256-dimensional feature vector on the $H/16 \times W/16$ feature pixel plane describes a specific region of the environment, while the $256$ channels represent different aspects of that environmental information. 
To optimize the transmission of the feature map over an $N \times N$ MIMO system, a mapping scheme from the feature map to $N$ data streams is required.
 
While traditional diversity and interleaving schemes rely on handcrafted redundancy and bit-level permutations to ensure reliability, these principles can be transcended for task-driven systems. Guided by the MIMO channel information, ANN-based encoder can adaptively distribute features of all spatial regions and aspects across different symbols and data streams of the MIMO system.

Motivated by this, MIMO channel-aware encoding is performed to enhance feature transmission fidelity over fading MIMO channels. The detailed method is illustrated in the Fig.~\ref{MIMO_encoding}. The backbone of the module is constructed as a fully convolutional network, including alternating $1\times 1$ and $3\times 3$ convolutional layers, along with GELU and BN. A shortcut branch is utilized to facilitate gradient propagation and improve train performance. 

To achieve flexible and adaptive feature encoding for MIMO fading channels, a joint attention module that integrates image information and MIMO channel information is designed. Specifically, the output of the first convolutional layer is processed by a global pooling block to represent image feature. Meanwhile, the equivalent SNR of each sub-channel, as shown in Eq.~(\ref{Eq_SNR}) is embedded into a feature space of $256/N$ dimension via a multilayer perceptron (MLP) block and then concatenated to a $256\times1\times1$ wireless channel feature. Using FC, GELU and sigmoid activation functions, the image feature and wireless channel feature are jointly processed to a $256\times1\times1$ attention weight ranging from [0,1], which is then applied to different feature channels. Since the proposed MIMO channel-aware encoding module does not change the feature dimension, $N_m$ identical modules can be stacked to enhance the model's capabilities, and the decoding module at the receiver employs the same structure. Through this process, the input feature map $P$ is transformed into a robust feature representation, denoted as $P_m$. 

Due to such a multi-layer convolutional architecture with the explicit embedding of the sub-channel SNR information, each individual feature element in $P_m$ is a weighted combination of all feature channels within a corresponding spatial neighborhood of $P$. 
This strategy enables the ANN to leverage high-quality sub-channels to protect important features, while simultaneously avoiding the severe distortion of critical information on degraded sub-channels. 
 Experimental results further demonstrate that this design significantly enhances feature transmission and instance segmentation quality under varying channel conditions.
 
\subsection{Feature Adjustment and MIMO Baseband Processing}\label{sec:baseband}
After hierarchical feature fusion and MIMO channel-aware encoding, the transmitter has achieved an SoM-feature, denoted as $P_m$, which is characterized by low data redundancy, strong task utility, and high MIMO channel robustness. 
To enable the efficient and reliable transmission of the SoM-feature over standard MIMO digital communication system, subsequent feature adjustment and MIMO baseband processing are required. 
Firstly, an FC layer is utilized to adjust the channel number of the SoM-feature to $C$. The parameter $C$ can be treated as a hyper-parameter of the NN, enabling flexible adjustment of the data compression ratio (CR). 
Given the previous hierarchical feature fusion and MIMO channel-aware encoding, $P_m$ has already achieved a well-represented form with high information density and high MIMO transmission robustness. Consequently, further compression is realized through a simple FC layer. Through end-to-end cascaded training of all modules, the FC layer learns to work in concert with the preceding encoder, achieving further channel compression while preserving the aforementioned properties of SoM-feature.

The compressed SoM-feature $\bbP_c$ with a dimension of $C\times H/16\times W/16$ is then mapped to the MIMO system's $N$ data streams. 
To achieve this, the $C$ feature channels are uniformly and sequentially divided into $N$ parts along the feature channel dimension. Each part is subsequently flattened to form data stream, resulting in a final output dimension of $N\times L$, where $L=CHW/256N$. 

Then the SoM-feature is passed through a nonlinear activation function for distribution transformation and the tanh function is selected. The tanh function constrains the values of the SoM-feature within a zero-centered range of [-1,1], which facilitates feature quantization and model convergence for digital transmission.
Additionally, outlier feature values caused by high bit error rate (BER) may lead to abnormal loss values and cause gradient fluctuation during backpropagation. However, the tanh function exhibits gradient saturation for extremely large or small inputs, which enhances the stability of training under transmission bit errors.

For the MIMO baseband processing module, we adopt a standard pipeline to ensure the seamless integration of the SoM-feature transmission scheme with conventional digital MIMO communication systems. Specifically, the feature is quantized with $m$-bits precision and modulated into QPSK symbols. Then, SVD precoding, as defined in Eq.~(\ref{trans_precoding}), is applied to generate the transmit symbols $\bbX$. The CR of the feature can be defined as:

\begin{equation}
\begin{aligned}
\text{CR}=\frac{m\times C\times \frac{H}{16}\times \frac{W}{16}}{8\times 3\times H\times W}=\frac{mC}{6144}.
\end{aligned}
\label{CR}
\end{equation}

At the receiver, the corresponding operations are sequentially performed: MIMO combining as Eq.~(\ref{recei_precoding}), channel equalization as Eq.~(\ref{equlization}), QPSK demodulation with hard decision, and $m$-bits quantization recovery. The obtained feature with a dimension of $N\times L$ can be reshaped to reconstruct the $C\times H/16\times W/16$ received feature $\bbP_{c}^{'}$. 
Due to the non-differentiable operations as quantization in the MIMO baseband processing, direct gradient backpropagation is not feasible. Therefore, we define the equivalent noise $\bbW$ to model the impact of MIMO digital transmission as:

\begin{equation}
\begin{aligned}
\bbW=(\bbP_{c}^{'}-\bbP_{C})_{\mathrm{detach}},
\end{aligned}
\label{detach}
\end{equation}
where the detach operation separates $\bbW$ from the computational graph of the ANN to exclude its gradient from the optimization. Therefore, we can model the impact of quantization and bit errors as:
\begin{equation}
\begin{aligned}
\bbP_{c}^{'}=\bbP_{c}+\bbW,
\end{aligned}
\label{noise_eq}
\end{equation}
and when computing the gradient of the loss funcition $L$ during the backpropagation, we have:
\begin{equation}
\begin{aligned}
\frac{\partial L}{\partial \bbP_{c}^{'}}=\frac{\partial L}{\partial \bbP_{c}}.
\end{aligned}
\label{gradient}
\end{equation}

In such a manner, the impact of MIMO digital transmission is modeled as a perturbation on the SoM-feature $\bbP_c$, where the distribution of $\bbW$ is jointly influenced by quantization errors and BER. After receiving $\bbP_c^{'}$, an FC layer is utilized to decompress it to $\bbP_m^{'}$ with 256 channels.

\subsection{Training Loss Function}\label{sec:loss}

As discussed in \cite{he2017mask} and \cite{ren2016faster}, for training the Mask R-CNN model on instance segmentation task, five losses are utilized, which includes RPN classification loss $L_{\mathrm{rpn\_cls}}$ and RPN bounding box regression loss $L_{rpn\_box}$ for training RPN as well as classification loss $L_{\mathrm{cls}}$, bounding box regression loss $L_{\mathrm{box}}$ and mask loss $L_{\mathrm{mask}}$ for training Mask R-CNN head. We retain these losses to enable end-to-end optimization of the instance segmentation task under MIMO feature transmission. Furthermore, to ensure the transmission fidelity of the SoM-feature, we introduce an additional mean squared error (MSE) loss between the feature pyramids at the transmitter and receiver as:
\begin{equation}
\begin{aligned}
L_{\mathrm{mse}}=\sum_{i=2}^{6}\text{MSE}(\bbP_{i}^{'},\bbP_{i}).
\end{aligned}
\label{mse}
\end{equation}

Through a hyperparameter $\lambda$, two parts of loss functions are weighted combined for ANN training as:

\begin{equation}
\begin{aligned}
L=L_{\mathrm{perception}}+\lambda L_{\mathrm{mse}},
\end{aligned}
\label{mse}
\end{equation}
where $L_{\mathrm{perception}}=L_{\mathrm{rpn\_cls}}+L_{\mathrm{rpn\_box}}+L_{\mathrm{cls}}+L_{\mathrm{box}}+L_{\mathrm{mask}}$. Due to the task-driven property of the the proposed scheme, $\lambda$ typically takes a small value, ensuring that $L_{\mathrm{mse}}$ serves as a regularization constraint to maintain transmission quality under adverse communication conditions.

\section{Experimental Results}\label{sec:simu}
\begin{table}[!t]
\normalsize
\centering
\caption{Training Settings.} 
\begin{tabular}{|c|c|}
\hline
\textbf{Parameters} & \textbf{Values}\\
\hline
Batch Size & $4$\\
\hline
Total Epoch & $50$\\
\hline
Warmup Epoch & $2.5$\\
\hline
Initial Learning Rate & $3e^{-4}$  \\
\hline
Final Learning Rate & $1e^{-7}$ \\
\hline
Weight Decay & $0.01$ \\
\hline
\end{tabular}
\label{tab:parameter_train}
\vspace{-4mm}
\end{table}

\begin{table}[!b]
\vspace{-4mm}
\normalsize
\centering
\caption{Model Settings.} 
\begin{tabular}{|c|c|}
\hline
\textbf{Parameters} & \textbf{Values}\\
\hline
$C$ & $\{24,48,96\}$\\
\hline
$m$ & $\{2,4,8\}$\\
\hline
$N_m$ & $6$\\
\hline
$N$ & $\{1,2,4\}$  \\
\hline
$\text{SNR}_{\mathrm{train}}$ & $[-5,15]$dB \\
\hline
$\text{SNR}_{\mathrm{test}}$ & $\{-5,0,5,10,15\}$dB \\
\hline
\end{tabular}
\label{tab:parameter_model}
\end{table}

To validate the proposed SoM-MIMO scheme, we train and evaluate our model on the widely used public urban traffic scenario dataset Cityscapes~\cite{cityscapes}, which contains 2975 training images and 500 validation images with a high resolution of $2048 \times 1024$. The mean average precision (mAP) is evaluated for instances across eight categories, including person, rider, car, truck, bus, train, motorcycle and bicycle. The instance-level mAP is computed as the mean of the average precision (AP) values across the eight classes, where AP for each class is calculated by averaging the AP values over ten different Intersection-over-Union (IoU) thresholds ranging from 0.5 to 0.95 in steps of 0.05. Further details of the datasets can be found in~\cite{cityscapes}.

The Mask R-CNN model utilized in our scheme is implemented based on the Detectron2~\cite{detectron2} platform, which provides state-of-the-art detection and segmentation algorithms. The Mask R-CNN model provided by the platform employs ResNet-50 as its backbone, which is pre-trained on the large-scale COCO~\cite{coco} dataset and subsequently trained on the Cityscapes dataset. 
The detailed training procedure of our proposed scheme is shown as follows:

Firstly, we initialize the backbone, FPN module and the Mask R-CNN head using the parameters from the Detectron2 model zoo. The parameters of the backbone and the FPN module are frozen in the training procedure as an optimal pre-trained feature extractor. Subsequently, the SoM-feature MIMO transmission modules are randomly initialized and cascaded with the task head for end-to-end training. This allows the training of the proposed SoM-MIMO system to focus on learning an efficient and robust MIMO transmission strategy and the corresponding feature utilization strategy. We employ the AdamW optimizer with a linear warmup cosine decay learning rate scheduler. The detailed training settings are provided in Table~\ref{tab:parameter_train}.
\begin{figure}[!t]
\centering{}
\includegraphics[width=0.50\textwidth]{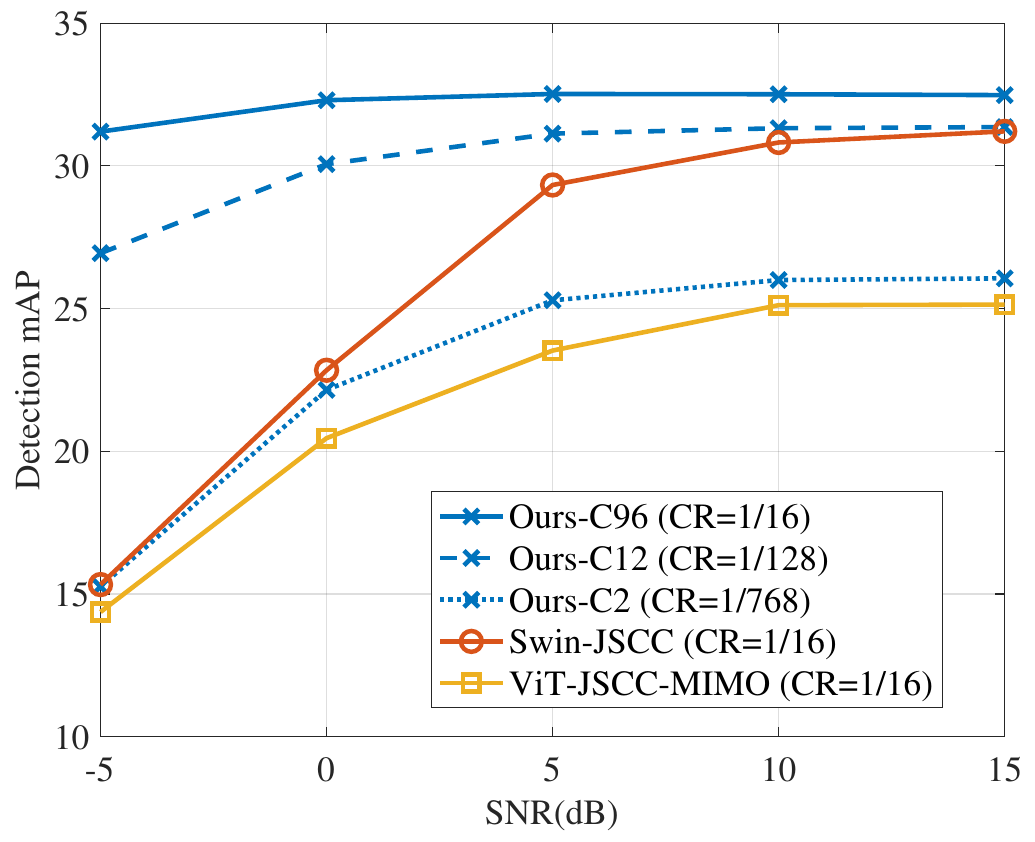}
\caption{Performance comparison between the proposed SoM-MIMO scheme and baseline schemes under $N=2$ and $m=4$.}{}
\label{Baseline comparison}
\end{figure}
\renewcommand{\arraystretch}{1.3} 
\begin{table}[!t]
\normalsize
\centering
\caption{Image PSNR of Baseline Schemes Under $N=2$, $m=4$, and $\text{CR}=1/16$.} 
\resizebox{0.49\textwidth}{!}
{
\begin{tabular}{|c|c|c|c|c|c|}
\hline
\multirow{2}{*}{\textbf{Scheme}} & \multicolumn{5}{|c|}{\textbf{SNR~(dB)}}\\ \cline{2-6}
& -5 & 0 & 5 & 10 & 15 \\ \hline
Swin-JSCC & $30.58$ & $33.56$ & $36.01$ & $37.21$ & $37.71$\\ \hline
ViT-JSCC-MIMO & $29.12$ & $31.49$ & $32.98 $ & $33.36$ & $33.40$\\ \hline 
\end{tabular}
}
\label{tab:PSNR}
\vspace{-4mm}
\end{table}

In terms of model settings, we choose $C=\{24,48,96\}$ for the SoM-feature channels and $m={2,4,8}$ for quantization precision, enabling adjustable CR. The depth of the MIMO channel-aware encoding and decoding module is set to $N_m=6$, and the weight of the MSE loss is assigned as $\lambda=0.1$. For the MIMO system, the number of antennas can be set to $N={1,2,4}$, and the channel is modeled as a power-normalized Rayleigh fading channel, as described in Sec.~\ref{seg:precoding}. During training, the $\text{SNR}_{\mathrm{train}}$ is uniformly distributed over the range of $[-5,15]$dB, allowing the model to comprehensively learn transmission strategies under diverse channel conditions. While during test, the $\text{SNR}_{\mathrm{test}}$ is set to $\{-5,0,5,10,15\}$dB, respectively. The relevant model settings are summarized in Table~\ref{tab:parameter_model}.

We compare our SoM-MIMO scheme with a Swin-Transformer-based DeepJSCC scheme~\cite{witt} (marked as Swin-JSCC), which is not specifically optimized for MIMO systems, and a ViT-based DeepJSCC scheme~\cite{closeloop-deniz}(marked as ViT-JSCC-MIMO), which is designed for MIMO systems with SVD precoding. Since both schemes are designed for image transmission, we train and evaluate the models on the Cityscapes dataset, using the PSNR as the training loss. To enable the ViT model to process high-resolution images with moderate computational overhead, we divide the images into $256 \times 256$ patches during training and testing the ViT-JSCC-MIMO scheme. 

\subsection{Baseline Schemes Comparison}\label{sec:baseline}

\begin{figure*}[!t]
    \centering
    \subfloat[Original Image]{
        \includegraphics[width=0.45\textwidth]{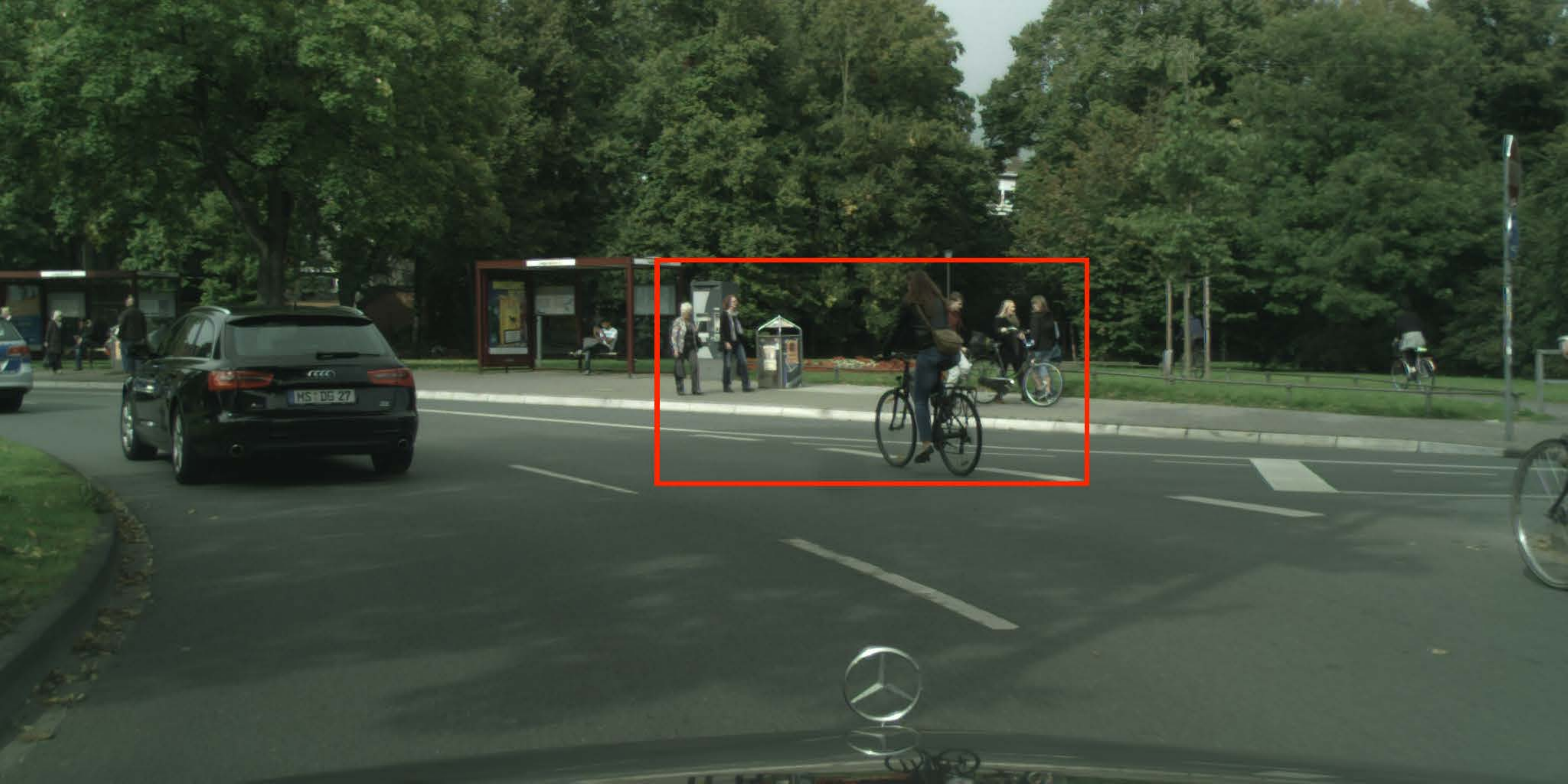}
        \label{fig:sub1}
    }
    \hfill
    \subfloat[Ground Truth]{
        \includegraphics[width=0.4\textwidth]{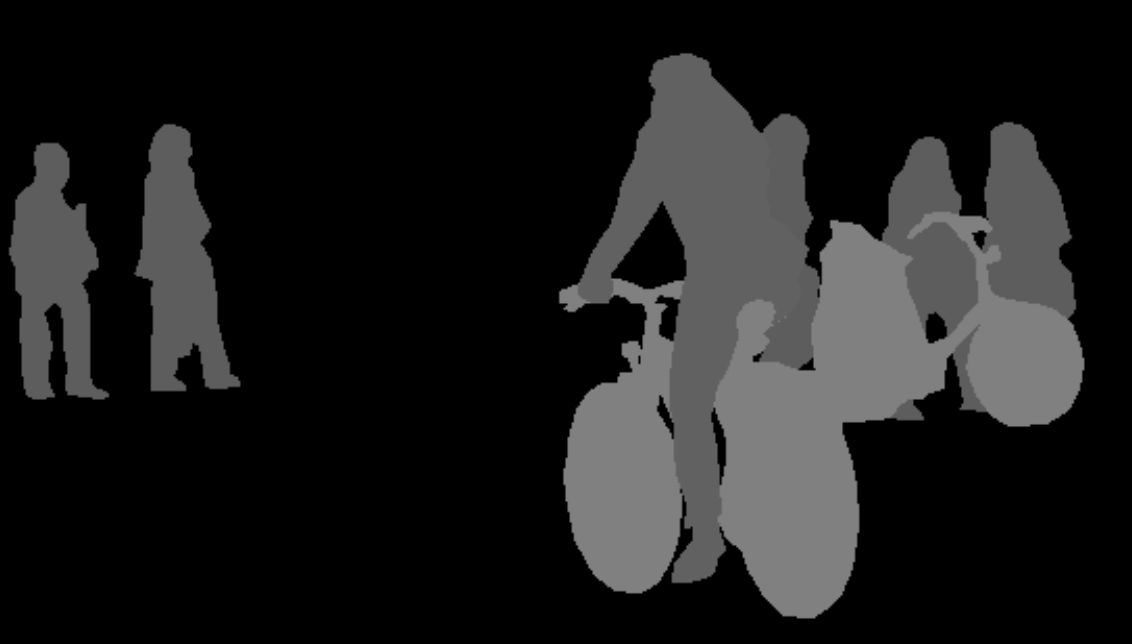}
        \label{fig:sub2}
    }
    \vspace{3pt} % 调整垂直间距
    
    \subfloat[Ours, 15dB]{
        \includegraphics[width=0.3\textwidth]{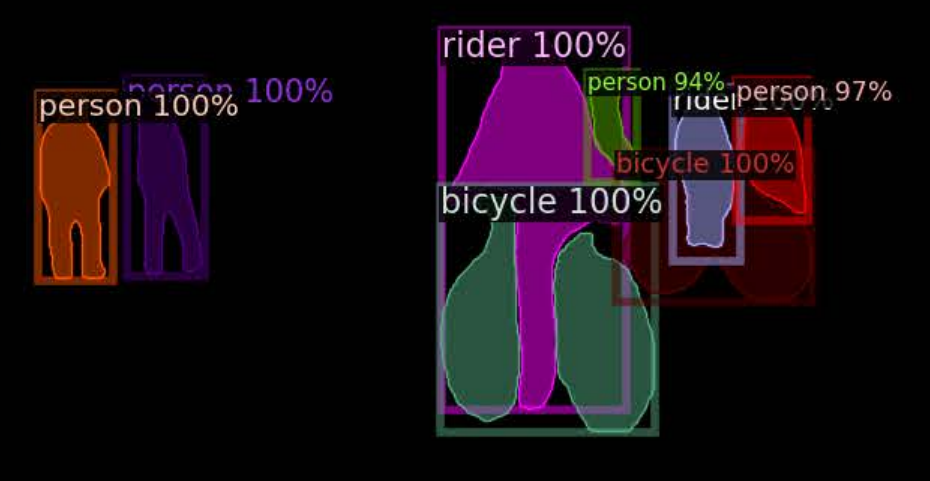}
        \label{fig:sub3}
    }
    \hfill
    \subfloat[Swin-JSCC, 15dB]{
        \includegraphics[width=0.3\textwidth]{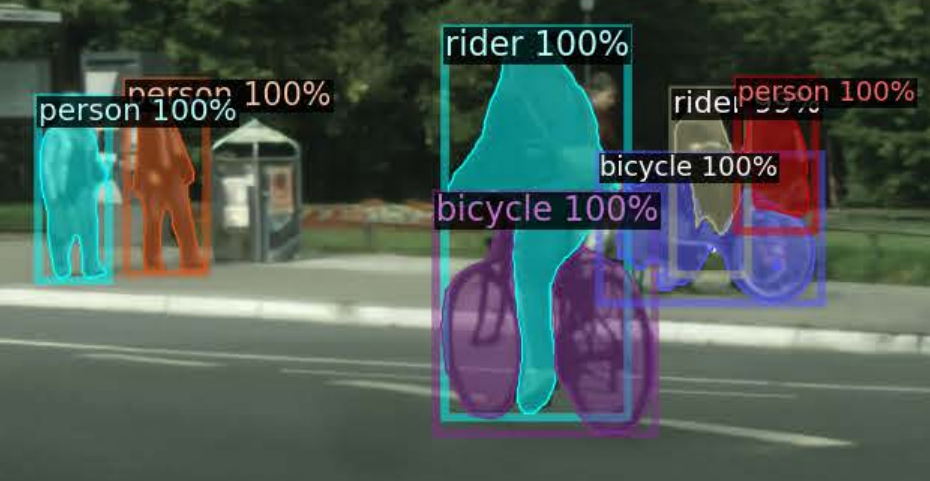}
        \label{fig:sub4}
    }
    \hfill
    \subfloat[ViT-JSCC-MIMO, 15dB]{
        \includegraphics[width=0.3\textwidth]{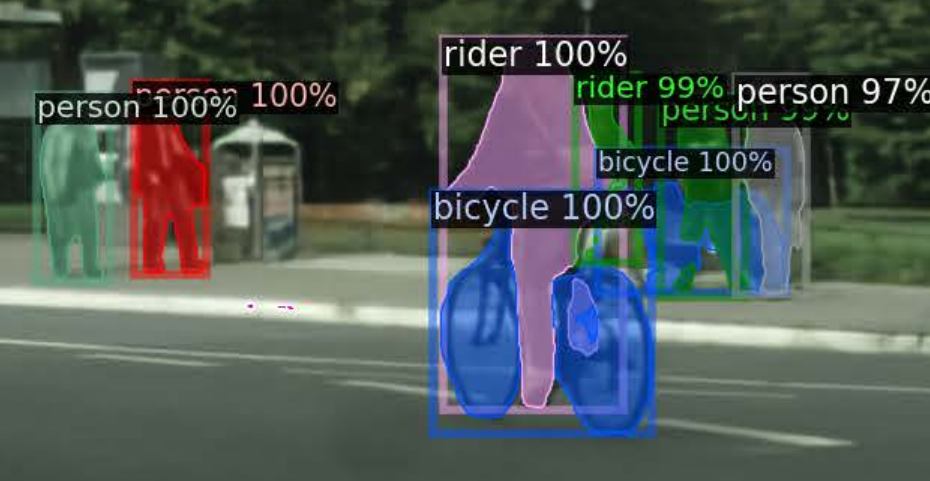}
        \label{fig:sub5}
    }

 \vspace{3pt} % 调整垂直间距
    \subfloat[Ours, -5dB]{
        \includegraphics[width=0.3\textwidth]{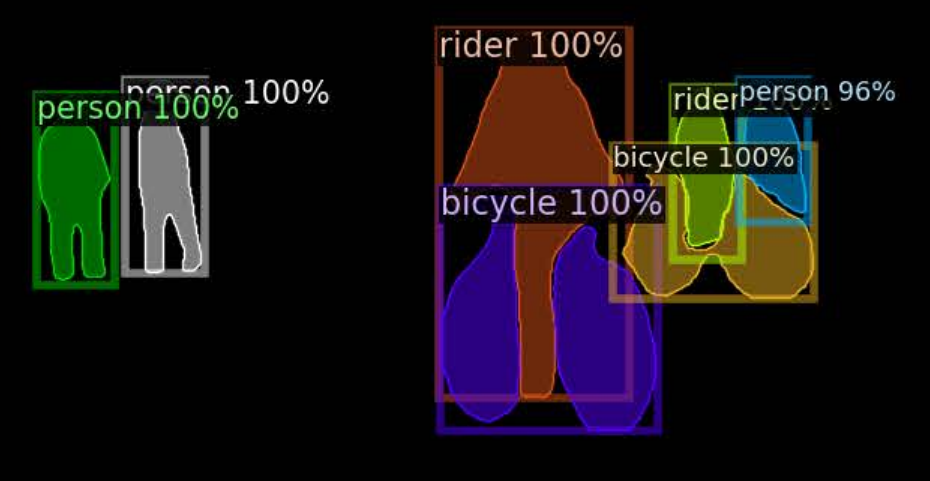}
        \label{fig:sub6}
    }
    \hfill
    \subfloat[Swin-JSCC, -5dB]{
        \includegraphics[width=0.3\textwidth]{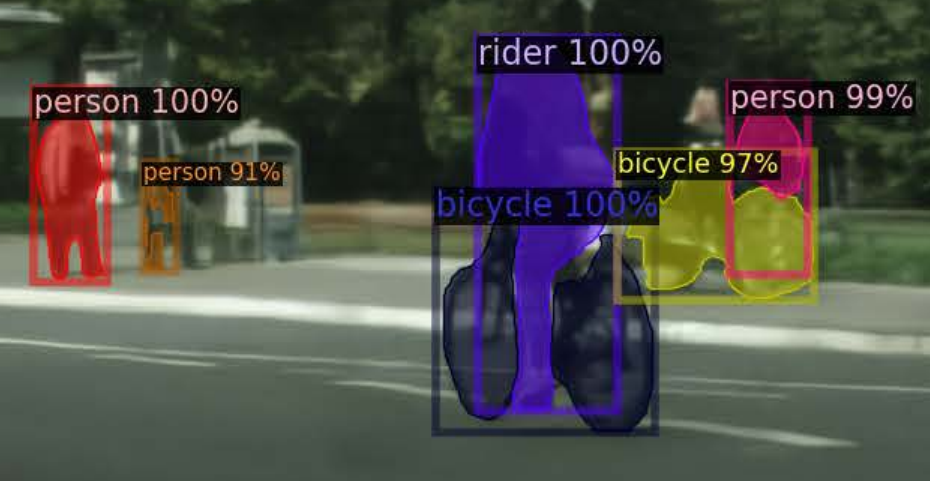}
        \label{fig:sub7}
    }
    \hfill
    \subfloat[ViT-JSCC-MIMO, -5dB]{
        \includegraphics[width=0.3\textwidth]{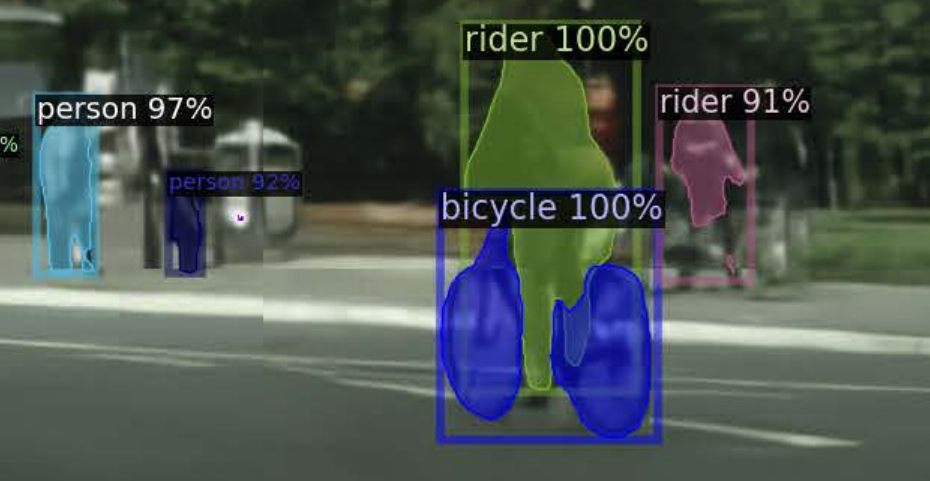}
        \label{fig:sub8}
    }
    \caption{Visualized performance comparison of our proposed scheme with Swin-JSCC and ViT-JSCC-MIMO under $CR=1/16$, $N=2$, and $m=4$.}
    \label{fig:visualization}
\end{figure*}

To ensure the fairness and validity of the baseline comparison, we established the following communication settings, network architectures, and training configurations: For communication settings, we replace the original DTAT strategy in both schemes with the same MIMO digital baseband processing strategy used in our scheme, which includes tanh activation, feature quantization, QPSK modulation and SVD precoding as detailed in Sec.~\ref{sec:baseband}. Additionally, during both training and testing, the same channel model and SNR settings as ours are employed. 

For network architectures, both schemes are implemented following the network architecture described in the original papers. The CR for both our scheme and the baseline schemes is set to $1/16$ as defined in Eq.~(\ref{CR}). 
Moreover, considering the difference in task between our scheme and the baseline schemes, we take the images reconstructed by the baseline schemes and feed them into a Mask R-CNN model. 
Specifically, we utilize the same backbone, FPN, RPN, and Mask R-CNN head parameters as those in our proposed scheme, just remove other modules that related to SoM-feature transmission.

For training configurations, both baseline schemes are firstly trained on the Cityscapes dataset to achieve reasonable PSNR performance. Subsequently, to ensure a fair task-level comparison, we use the images reconstructed by these schemes to train the Mask R-CNN model. Similarly, the backbone and FPN are frozen, while RPN and head are training according to the settings shown in Table~\ref{tab:parameter_train}.

To summarize, we conduct a fair comparison with the baseline schemes under identical experimental conditions, including the dataset, task, MIMO transmission condition, and instance segmentation model capabilities. The PSNR performance of the baseline schemes is shown in Table~\ref{tab:PSNR} and the mAP performance comparison is illustrated in the Fig.~\ref{Baseline comparison}. When the SNR decreases, higher BER leads to increased blur and noise in the reconstructed images, resulting in a decline in the PSNR. For instance segmentation tasks in complex traffic scenarios, environmental images contain dense small objects, such as distant pedestrians and bicycles, making mAP performance highly sensitive to blur and noise. This issue results in lower robustness to bit errors for baseline schemes that separate image transmission and instance segmentation. Moreover, due to the limitations of the ViT model in processing dense information in high-resolution images~\cite{liu2021swin}, the performance of the ViT-JSCC-MIMO schemes on the Cityscapes dataset is inferior to that of the Swin-JSCC scheme. 

However, evidenced by the experimental results, our proposed scheme shows higher instance segmentation performance and stronger robustness to bit errors under the same communication overhead as the baseline schemes, achieving average mAP improvements of 6.30 and 10.48 across all SNR levels over two baseline schemes, as well as 15.87 and 16.83 improvements at -5dB SNR. Moreover, when achieving performance comparable to the baseline schemes, our proposed scheme significantly reduces transmission overhead. Specifically, when $\text{CR}=1/128$, we have performance comparable to Swin-JSCC at high SNR levels, while at $\text{CR}=1/768$, it attains performance comparable to the two baseline schemes at low SNR levels and performance comparable to ViT-JSCC-MIMO at high SNR levels. This provides significant flexibility in balancing communication overhead and perception performance.

\begin{figure}[!t]
\centering{}
\includegraphics[width=0.50\textwidth]{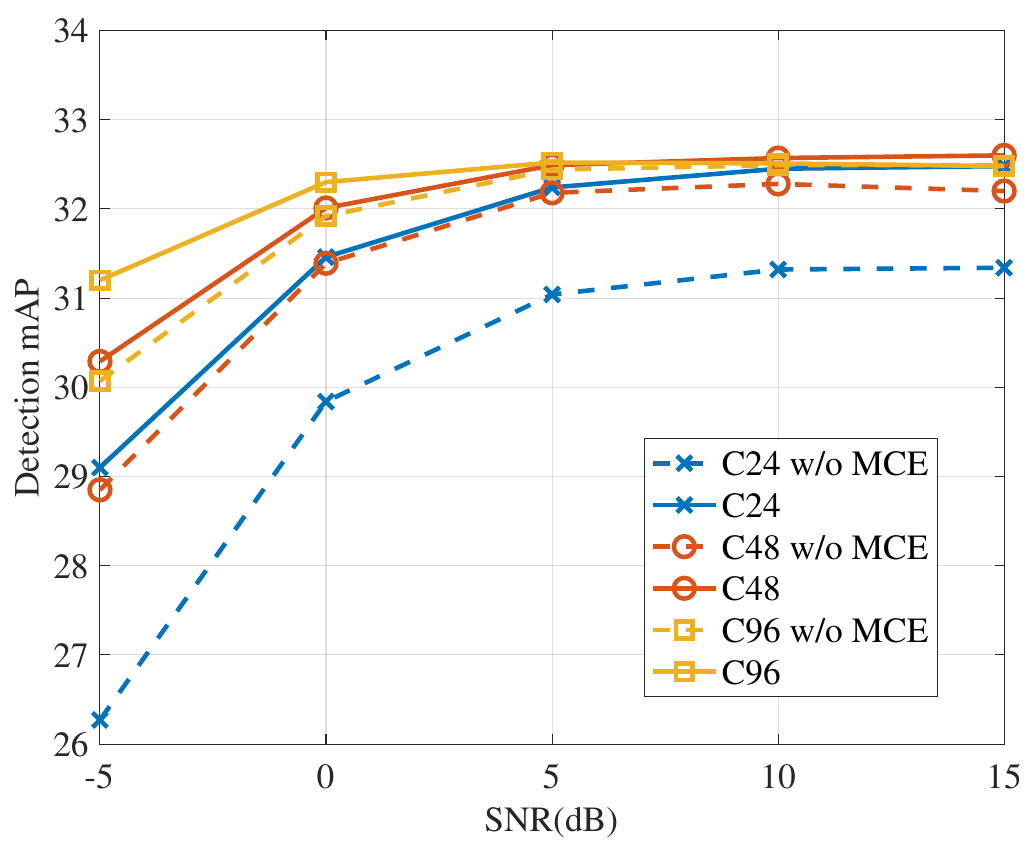}
\caption{Ablation study on the MIMO channel-aware encoding and decoding modules under $N=2$ and $m=4$ with $C=\{24,28,96\}$.}{}
\label{ablation}
\vspace{-4mm}
\end{figure}

In the visualized example shown in Fig.~\ref{fig:visualization}, we compare the instance segmentation performance in a complex traffic scene containing four pedestrians, two riders, and two bicycles, with a pedestrian significantly occluded. 
Our proposed method maintain reliable instance segmentation performance at both 15dB and -5dB, with only a missed detection of the occluded pedestrian at -5dB. However, both baseline schemes exhibit severe missed detections and false detections due to image blurring and distortion, especially at -5dB.

As evidenced by both quantitative and visualized comparisons, our proposed scheme can achieve superior perception performance with reduced communication overhead under adverse communication conditions by integrating the instance segmentation task with MIMO digital transmission. Subsequent experiments will further validate the effectiveness of MIMO channel information embedding in our scheme and the universality of our proposed scheme across different MIMO system settings.

\subsection{Ablation Study on MIMO Channel-aware Encoding and Decoding Modules}\label{sec:abation}

We conducted an ablation study to validate the effectiveness of our proposed MIMO channel-aware encoding and decoding modules. By removing the MCE and MCD modules, feature $\bbP$ from the HFF module is directly fed into the FC layer for channel compression at the transmitter, while the feature $\bbP^{'}_{m}$ from the FC layer after channel decompression are sent to HFS module at the receiver. In the ablated model, the MIMO fading channel is not actively learned through explicit utilization of the channel information but is passively adapted during the end-to-end training phase. The result of the ablation study is shown in Fig.~\ref{ablation}. As mentioned in Table~\ref{tab:parameter_model}, we train the model with $\text{SNR}_{train}$ uniformly distributed in $[-5,15]$dB, and test it at multiple fixed SNR levels, which imposes high demands on the model's adaptability to varying SNR conditions.
\begin{figure}[!t]
\centering{}
\includegraphics[width=0.50\textwidth]{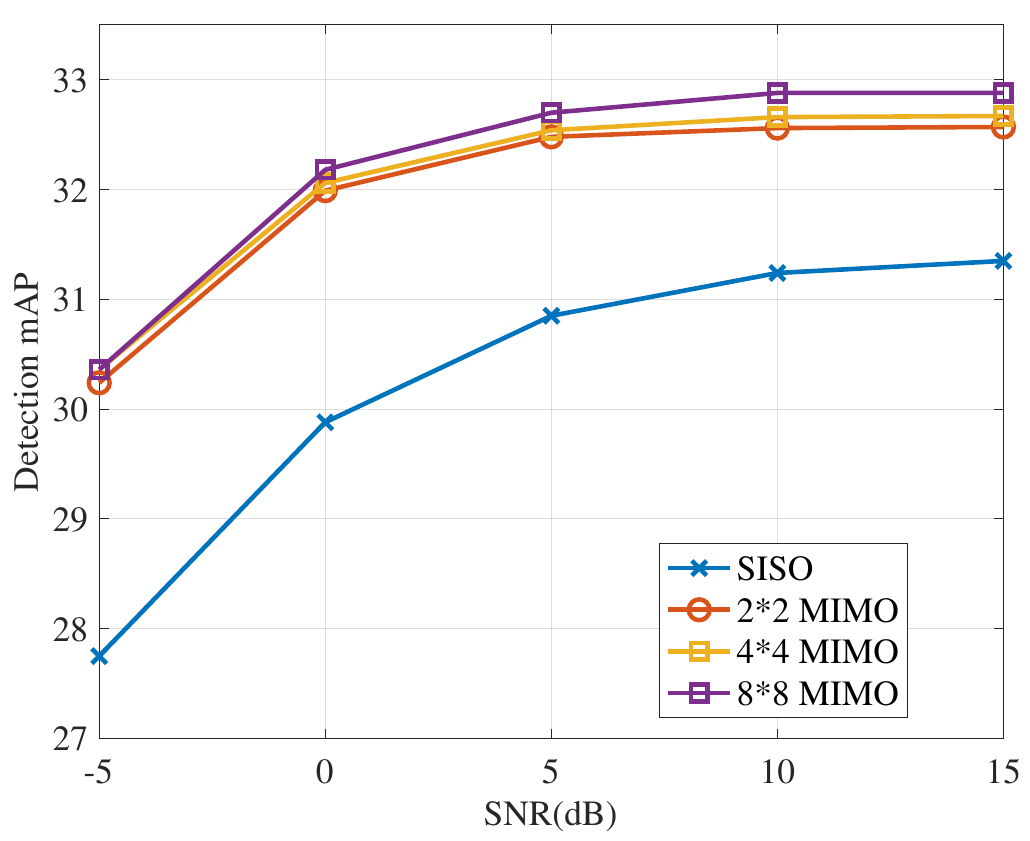}
\caption{Performance comparison with different antenna numbers under $N=\{1,2,4\}$, $C=48$ and $m=4$.}{}
\label{antenna}
\vspace{-4mm}
\end{figure}
The experimental results demonstrate that our proposed $\text{MCE}$ and $\text{MCD}$ modules significantly improve the performance across different compressed SoM-feature channel numbers $C$. Specifically, the ablated scheme fails to effectively capture the appropriate feature transmission strategy, which leads to both performance saturation at high $\text{SNR}$s and performance degradation at low $\text{SNR}$s. 
Furthermore, it is observed that the performance gap becomes larger as $C$ decreases, highlighting the increased importance of embedding the MIMO transmission model into the neural network to guide feature encoding when the redundancy of transmitted information is reduced.
However, by leveraging the MCE and MCD modules, our proposed method achieves generalizability across diverse MIMO communication conditions through a single training procedure with mixed SNR levels, which enhances the efficiency of both training and deployment.
\renewcommand{\arraystretch}{1.3} 
\begin{table*}[!t]
\centering
\caption{Sub-channel BER Under Different MIMO System Settings}
\begin{tabular}{|c|c|c|c|c|c|c|}
\hline
\multirow{2}{*}{\textbf{N}}& \multirow{2}{*}{\textbf{Stream}} & \multicolumn{5}{|c|}{\textbf{SNR~(dB)}} \\ \cline{3-7}                                                              
& & -5 & 0 & 5 & 10 & 15 \\ \hline
1 & \textbf{1} & $\mathbf{3.15\times 10^{-1}}$ & $\mathbf{2.13\times 10^{-1}}$ & $\mathbf{1.10\times 10^{-1}}$ & $\mathbf{4.47\times 10^{-2}}$ & $\mathbf{1.43\times 10^{-2}}$ \\ \hline
\multirow{3}{*}{2} & 1 & $2.39\times 10^{-1}$ & $1.12\times 10^{-1}$ & $2.38\times 10^{-2}$ & $1.73\times 10^{-3}$ & $4.46\times 10^{-5}$ \\ \cline{2-7}
 & 2 & $4.02\times 10^{-1}$ & $3.34\times 10^{-1}$ & $2.33\times 10^{-1}$ & $1.31\times 10^{-1}$ & $5.43\times 10^{-2}$ \\ \cline{2-7}
 & \textbf{Avg} & $\mathbf{3.21\times 10^{-1}}$ & $\mathbf{2.23\times 10^{-1}}$ & $\mathbf{1.28\times 10^{-1}}$ & $\mathbf{6.63\times 10^{-2}}$ & $\mathbf{2.72\times 10^{-2}}$ \\ \hline
 \multirow{5}{*}{4} & 1 & $1.94\times 10^{-1}$ & $6.52\times 10^{-2}$ & $5.03\times 10^{-3}$ & $1.99\times 10^{-5}$ & $\textless 1 \times 10^{-7}$ \\ \cline{2-7}
 & 2 & $2.81\times 10^{-1}$ & $1.54\times 10^{-1}$ & $3.96\times 10^{-2}$ & $2.02\times 10^{-3}$ & $7.67 \times 10^{-6}$ \\ \cline{2-7}
 & 3 & $3.67\times 10^{-1}$ & $2.73\times 10^{-1}$ & $1.48\times 10^{-1}$ & $4.12\times 10^{-2}$ & $3.97 \times 10^{-3}$ \\ \cline{2-7}
 & 4 & $4.50\times 10^{-1}$ & $4.13\times 10^{-1}$ & $3.50\times 10^{-1}$ & $2.57\times 10^{-1}$ & $1.48 \times 10^{-1}$ \\ \cline{2-7}
 & \textbf{Avg} & $\mathbf{3.23\times 10^{-1}}$ & $\mathbf{2.26\times 10^{-1}}$ & $\mathbf{1.35\times 10^{-1}}$ & $\mathbf{7.51\times 10^{-2}}$ & $\mathbf{3.81 \times 10^{-2}}$ \\ \hline
\end{tabular}
\label{tab:BER}
\vspace{-4mm}
\end{table*}
\subsection{Impact of Antenna Number on Performance}\label{sec:antenna}

To validate the scalability of our proposed scheme with respect to the number of antennas, we conducted a performance comparison of our proposed scheme under $N=\{1,2,4,8\}$ while maintaining the same CR. Following the module design described in Sec.~\ref{sec:mimo}, our MCE and MCD modules can flexibly adapt to different numbers of antennas by simply adjusting the dimensions of the MLP used for embedding channel information. 

For MIMO systems employing SVD precoding, both transmission efficiency and average BER increase with the number of antennas. However, due to the descending order of the singular value $\lambda_1,\cdots,\lambda_N$, the first few data streams in MIMO systems exhibit lower BER, as shown in Table.~\ref{tab:BER}. Experimental results as shown in Fig.~\ref{antenna} indicate that, compared to SISO systems, the performance of the SoM-feature MIMO transmission network can be significantly improved by adaptively encoding features into multiple sub-channels of varying quality. Specifically, during both training and testing, the proposed method can leverage data streams with lower BER to effectively counteract the impact of fading channel and noise on perception performance. As the number of antennas continues to increase, the BER of the first few data streams decreases, while the difficulty of appropriately encoding SoM-features into more sub-channels grows, which increases the learning difficulty of the model. As a result, the improvement in SoM-feature transmission performance tends to decline significantly.

\subsection{Impact of Quantization Precision on Performance}\label{sec:quant}
\begin{figure}[!b]
\centering{}
\includegraphics[width=0.50\textwidth]{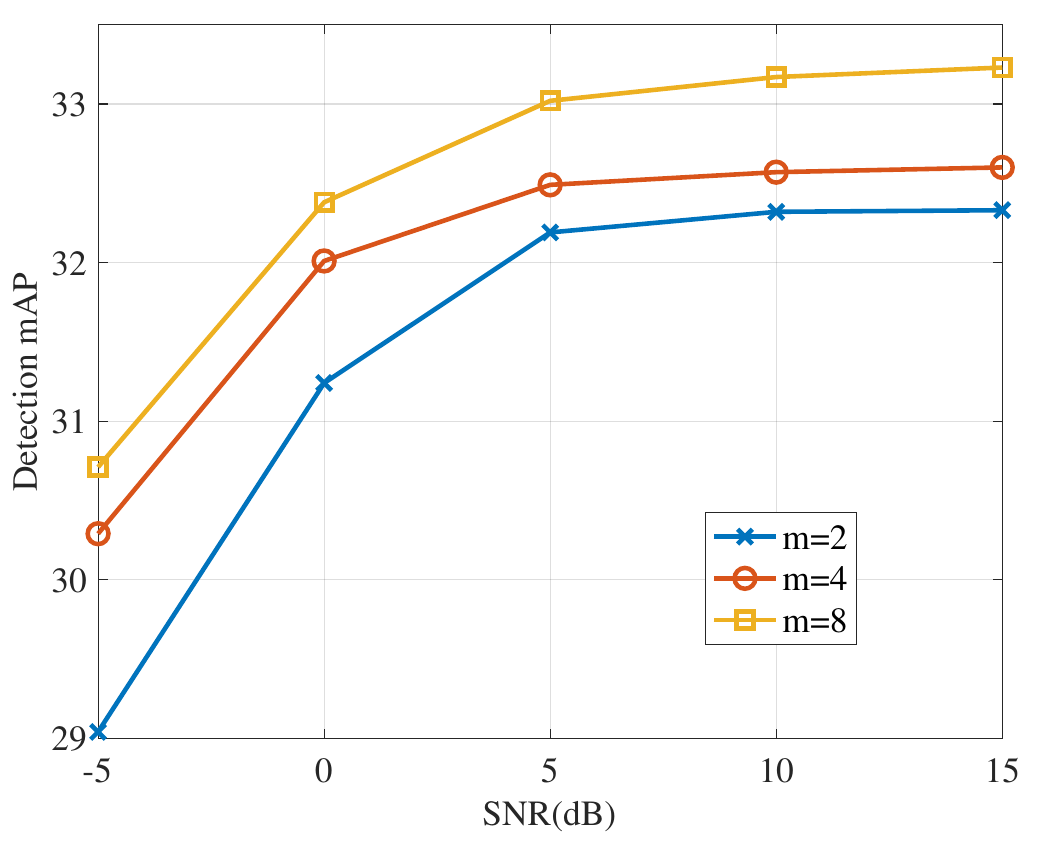}
\caption{Performance comparison with different quantization precision under $N=2$ and $C=48$.}{}
\label{bits}
\end{figure}
Since our proposed scheme employs standard quantization and QPSK modulation in the MIMO baseband processing, it can also flexibly adapt to different quantization precision $m$. The reduction in $m$ primarily affects the SoM-feature transmission in two aspects: firstly, reduced quantization precision results in higher quantization errors; secondly, as the influence of each bit on the feature values increases, the robustness of the features to bit errors decreases. This leads to an increase in $\bbW$ in Eq.~(\ref{noise_eq}), which may potentially introduce perturbations to the model. However, as shown in Fig.~\ref{bits}, thanks to the task-oriented design, MIMO channel embedding modules, and feature adjustment for digital transmission, our proposed scheme demonstrates robust performance across different quantization precision levels. This provides our proposed scheme with enhanced flexibility in adjusting communication overhead.

In summary, the experimental results demonstrate that, compared to JSCC schemes designed for image transmission, our proposed scheme achieves significantly superior instance segmentation performance with lower communication cost under adverse MIMO communication conditions. 
Additionally, the proposed MIMO channel-aware encoding and decoding modules effectively enhance the model's adaptability to varying MIMO communication conditions. 
Moreover, our scheme exhibits strong adaptability to different numbers of antennas and quantization levels, leading to greater flexibility for practical MIMO digital communication systems.

\section{Conclusions and Future Work}\label{sec:conclusion}
This paper presents an SoM-based task-driven MIMO system for image transmission: SoM-MIMO. Compared with the existing work on MIMO JSCC, our scheme tightly integrates the instance segmentation task with the digital MIMO communication systems. Specifically, the feature pyramid from the backbone is first hierarchically aggregated based on its structural properties, and then adaptively encoded by incorporating the MIMO channel information with SVD precoding to obtain compact and robust SoM-feature. Subsequently, standard MIMO baseband processing with nonlinear feature activation is applied to the SoM-feature to enable its transmission over the MIMO channel. This integrated design ensures the fidelity of critical task-related information while ensuring robustness under adverse communication conditions. Experimental results demonstrate that SoM-MIMO outperforms existing JSCC-based perception-communication decoupled schemes in instance segmentation accuracy, particularly under challenging channel conditions.

Based on the research in this paper, future research can explore the following areas: In real-world applications, there may be cooperation among multiple mobile agents, requiring one mobile agent to transmit perceptual information obtained from different regions or modalities to different cooperator. This presents the opportunity for joint design of multi-user communications and CP. Furthermore, mobile agents need to fuse the received SoM-feature with its own perceptual information, taking into account both the fidelity of the transmitted information and its contribution to the task. Therefore, it is necessary to design attention mechanisms based on feature quality and channel conditions to achieve flexible weighted fusion, fully leveraging the performance gains of CP.

\bibliographystyle{IEEEtran}
\bibliography{ref.bib}

\begin{thebibliography}{10}
\providecommand{\url}[1]{#1}
\csname url@samestyle\endcsname
\providecommand{\newblock}{\relax}
\providecommand{\bibinfo}[2]{#2}
\providecommand{\BIBentrySTDinterwordspacing}{\spaceskip=0pt\relax}
\providecommand{\BIBentryALTinterwordstretchfactor}{4}
\providecommand{\BIBentryALTinterwordspacing}{\spaceskip=\fontdimen2\font plus
\BIBentryALTinterwordstretchfactor\fontdimen3\font minus \fontdimen4\font\relax}
\providecommand{\BIBforeignlanguage}[2]{{%
\expandafter\ifx\csname l@#1\endcsname\relax
\typeout{** WARNING: IEEEtran.bst: No hyphenation pattern has been}%
\typeout{** loaded for the language `#1'. Using the pattern for}%
\typeout{** the default language instead.}%
\else
\language=\csname l@#1\endcsname
\fi
#2}}
\providecommand{\BIBdecl}{\relax}
\BIBdecl

\bibitem{classification}
A.~Krizhevsky, I.~Sutskever, and G.~E. Hinton, ``Imagenet classification with deep convolutional neural networks,'' in \emph{Proc. NIPS}, vol.~25, Lake Tahoe, USA, Dec. 2012.

\bibitem{yolo}
J.~Redmon, S.~Divvala, R.~Girshick, and A.~Farhadi, ``You only look once: Unified, real-time object detection,'' in \emph{Proc. CVPR}, Las Vegas, NV, USA, June 2016, pp. 779--788.

\bibitem{he2017mask}
K.~He, G.~Gkioxari, P.~Doll{\'a}r, and R.~Girshick, ``Mask {R-CNN},'' in \emph{Proc. ICCV}, Venice, Italy, Oct. 2017, pp. 2961--2969.

\bibitem{gridmap}
X.~Zheng, Y.~Li, D.~Duan, L.~Yang, C.~Chen, and X.~Cheng, ``Multivehicle multisensor occupancy grid maps ({MVMS-OGM}) for autonomous driving,'' \emph{IEEE Internet Things J.}, vol.~9, no.~22, pp. 22\,944--22\,957, Nov. 2022.

\bibitem{where2com}
Y.~Hu, S.~Fang, Z.~Lei, Y.~Zhong, and S.~Chen, ``Where2comm: Communication-efficient collaborative perception via spatial confidence maps,'' \emph{Proc. NIPS}, vol.~35, pp. 4874--4886, Nov. 2022.

\bibitem{confidence}
X.~Zheng, S.~Li, Y.~Li, D.~Duan, L.~Yang, and X.~Cheng, ``Confidence evaluation for machine learning schemes in vehicular sensor networks,'' \emph{IEEE Trans. Wireless Commun.}, vol.~22, no.~4, pp. 2833--2846, Apr. 2023.

\bibitem{ldpc}
T.~Richardson and S.~Kudekar, ``Design of low-density parity check codes for {5G} new radio,'' \emph{IEEE Commun. Mag.}, vol.~56, no.~3, pp. 28--34, Mar. 2018.

\bibitem{DJSCC}
E.~Bourtsoulatze, D.~Burth~Kurka, and D.~Gündüz, ``Deep joint source-channel coding for wireless image transmission,'' \emph{IEEE Trans. Cognit. Commun. Networking}, vol.~5, no.~3, pp. 567--579, May 2019.

\bibitem{SoM}
X.~Cheng, H.~Zhang, J.~Zhang, S.~Gao, S.~Li, Z.~Huang, L.~Bai, Z.~Yang, X.~Zheng, and L.~Yang, ``Intelligent multi-modal sensing-communication integration: Synesthesia of machines,'' \emph{IEEE Commun. Surv. Tutorials}, vol.~26, no.~1, pp. 256--301, Firstquarter 2024.

\bibitem{xu2021wireless}
J.~Xu, B.~Ai, W.~Chen, A.~Yang, P.~Sun, and M.~Rodrigues, ``Wireless image transmission using deep source channel coding with attention modules,'' \emph{IEEE Trans. Circuit Syst. Video Technol.}, vol.~32, no.~4, pp. 2315--2328, Apr. 2022.

\bibitem{witt}
K.~Yang, S.~Wang, J.~Dai, K.~Tan, K.~Niu, and P.~Zhang, ``{WITT}: A wireless image transmission transformer for semantic communications,'' in \emph{Proc. ICASSP}, Rhodes Island, Greece, Jun. 2023.

\bibitem{fast}
K.~Zhou, G.~Zhang, Y.~Cai, Q.~Hu, and G.~Yu, ``{FAST}: Feature arrangement for semantic transmission,'' in \emph{Proc. WCNC}, Dubai, United Arab Emirates, Apr. 2024.

\bibitem{early-mimo}
X.~Zhang, M.~Vaezi, and T.~J. O’Shea, ``{SVD}-embedded deep autoencoder for {MIMO} communications,'' in \emph{Proc. ICC}, Seoul, Korea, May 2022, pp. 5190--5195.

\bibitem{openloop-zp}
S.~Yao, S.~Wang, J.~Dai, and K.~Niu, ``Learned image transmission over {MIMO} fading channels,'' in \emph{Proc. PIMRC}, Toronto, ON, Canada, Sep. 2023.

\bibitem{openloop-csg}
B.~Xie, Y.~Wu, Y.~Shi, W.~Zhang, S.~Cui, and M.~Debbah, ``Robust image semantic coding with learnable {CSI} fusion masking over {MIMO} fading channels,'' \emph{IEEE Trans. Wireless Commun.}, vol.~23, no.~10, pp. 14\,155--14\,170, Oct. 2024.

\bibitem{closeloop-ab}
W.~Jiang, W.~Chen, and B.~Ai, ``Deep joint source channel coding with attention modules over {MIMO} channels,'' in \emph{Proc. VTC}, Singapore, Singapore, Jun. 2024.

\bibitem{closeloop-scan}
G.~Zhang, Q.~Hu, Y.~Cai, and G.~Yu, ``{SCAN}: Semantic communication with adaptive channel feedback,'' \emph{IEEE Trans. Cognit. Commun. Networking}, vol.~10, no.~5, pp. 1759--1773, Oct. 2024.

\bibitem{closeloop-deniz}
H.~Wu, Y.~Shao, C.~Bian, K.~Mikolajczyk, and D.~G{\"u}nd{\"u}z, ``Deep joint source-channel coding for adaptive image transmission over {MIMO} channels,'' \emph{IEEE Trans. Wireless Commun.}, vol.~23, no.~10, pp. 15\,002--15\,017, Oct. 2024.

\bibitem{closeloop-diffusion}
Y.~Duan, T.~Wu, Z.~Chen, and M.~Tao, ``{DM-MIMO}: Diffusion models for robust semantic communications over {MIMO} channels,'' in \emph{Proc. ICCC}, Aug. 2024, pp. 1609--1614.

\bibitem{digital-deniz}
T.-Y. Tung, D.~B. Kurka, M.~Jankowski, and D.~G{\"u}nd{\"u}z, ``{DeepJSCC-Q}: Constellation constrained deep joint source-channel coding,'' \emph{{IEEE} Trans. Inform. Theory}, vol.~3, no.~4, pp. 720--731, Dec 2022.

\bibitem{digital-tmx}
Y.~Bo, Y.~Duan, S.~Shao, and M.~Tao, ``Joint coding-modulation for digital semantic communications via variational autoencoder,'' \emph{IEEE Trans. Commun.}, vol.~72, no.~9, pp. 5626--5640, Sep. 2024.

\bibitem{digital-ab}
L.~Guo, W.~Chen, Y.~Sun, and B.~Ai, ``Device-edge digital semantic communication with trained non-linear quantization,'' in \emph{Proc. VTC}, Florence, Italy, Jun. 2023.

\bibitem{sdac}
Z.~Bao, C.~Dong, and X.~Xu, ``{sDAC}--semantic digital analog converter for semantic communications,'' \emph{arXiv preprint arXiv:2405.02335}, 2024.

\bibitem{task-class}
M.~Jankowski, D.~G{\"u}nd{\"u}z, and K.~Mikolajczyk, ``Wireless image retrieval at the edge,'' \emph{IEEE J. Sel. Areas Commun.}, vol.~39, no.~1, pp. 89--100, Jan. 2021.

\bibitem{task-segmentation}
Q.~Pan, H.~Tong, J.~Lv, T.~Luo, Z.~Zhang, C.~Yin, and J.~Li, ``Image segmentation semantic communication over internet of vehicles,'' in \emph{Proc. WCNC}, Glasgow, United Kingdom, Mar. 2023.

\bibitem{ren2016faster}
S.~Ren, K.~He, R.~Girshick, and J.~Sun, ``Faster {R-CNN}: Towards real-time object detection with region proposal networks,'' \emph{IEEE Trans. Pattern Anal. Mach. Intell.}, vol.~39, no.~6, pp. 1137--1149, Jun. 2017.

\bibitem{lin2017feature}
T.-Y. Lin, P.~Doll{\'a}r, R.~Girshick, K.~He, B.~Hariharan, and S.~Belongie, ``Feature pyramid networks for object detection,'' in \emph{Proc. CVPR}, Honolulu, HI, USA, Jul. 2017, pp. 2117--2125.

\bibitem{he2016deep}
K.~He, X.~Zhang, S.~Ren, and J.~Sun, ``Deep residual learning for image recognition,'' in \emph{Proc. CVPR}, Las Vegas, NV, USA, Jun. 2016, pp. 770--778.

\bibitem{liu2021swin}
Z.~Liu, Y.~Lin, Y.~Cao, H.~Hu, Y.~Wei, Z.~Zhang, S.~Lin, and B.~Guo, ``Swin transformer: Hierarchical vision transformer using shifted windows,'' in \emph{Proc. ICCV}, Venice, Italy, Oct. 2021, pp. 10\,012--10\,022.

\bibitem{senet}
J.~Hu, L.~Shen, and G.~Sun, ``Squeeze-and-excitation networks,'' in \emph{Proc. CVPR}, Salt Lake City, UT, USA, June 2018.

\bibitem{cityscapes}
M.~Cordts, M.~Omran, S.~Ramos, T.~Rehfeld, M.~Enzweiler, R.~Benenson, U.~Franke, S.~Roth, and B.~Schiele, ``The cityscapes dataset for semantic urban scene understanding,'' in \emph{Proc. CVPR}, Las Vegas, NV, USA, June 2016, pp. 3213--3223.

\bibitem{detectron2}
Y.~Wu, A.~Kirillov, F.~Massa, W.-Y. Lo, and R.~Girshick, ``Detectron2,'' \url{https://github.com/facebookresearch/detectron2}, 2019.

\bibitem{coco}
T.-Y. Lin, M.~Maire, S.~Belongie, J.~Hays, P.~Perona, D.~Ramanan, P.~Doll{\'a}r, and C.~L. Zitnick, ``Microsoft coco: Common objects in context,'' in \emph{Proc. ECCV}, Zurich, Switzerland, Sep. 2014, pp. 740--755.

\end{thebibliography}

\end{document}